\documentclass[submit]{smj}

\usepackage[english]{babel}
\usepackage{bm}
\usepackage[table]{xcolor}
\usepackage{multirow}
\usepackage{float}
\usepackage{rotating}
\usepackage{pdflscape}

\usepackage{etoolbox}
\usepackage{url}
\DeclareGraphicsExtensions{.pdf,.png,.jpg,.eps}

\makeatletter
\patchcmd{\@makecaption}
  {\parbox}
  {\advance\@tempdima-\fontdimen2} % decrease the width!
  {}{}
\makeatother  

\usepackage[caption=false]{subfig}
\usepackage{enumerate}
\usepackage{comment}
\usepackage{float}

\Author{Leonardo Egidi \Affil{1}, 
        Francesco Pauli\Affil{2}, 
        and Nicola Torelli\Affil{2}
}
\AuthorRunning{Egidi \textrm{et al.}}

\Affiliations{

\item Dipartimento di Scienze Statistiche,
      Universit\`a degli Studi di Padova,
      Padova, Italy

\item Dipartimento di Scienze Economiche, Aziendali, Matematiche e Statistiche `Bruno de Finetti',
      Universit\`{a} degli Studi di Trieste,
      Trieste, Italy

}   %% end \Affiliations

\CorrAddress{Leonardo Egidi, 
             Dipartimento di Scienze Statistiche,
      Universit\`a degli Studi di Padova,
      Padova,
             Via C. Battisti, 241 - 35121 Padova, 
             Italy}
\CorrEmail{egidi@stat.unipd.it}
\CorrPhone{(+39)\;049\;827\;4174}
\Title{Combining historical data and bookmakers'odds in modelling football scores}
\TitleRunning{Modelling football scores}

\Abstract{
Modelling football outcomes has gained increasing attention, in large part due to the potential for making substantial 
profits. Despite the strong connection existing between football models and the bookmakers' betting odds, no authors 
have used the latter for improving the fit and the predictive accuracy of these models. We have developed 
a hierarchical Bayesian Poisson model in which the scoring rates of the teams are convex combinations of parameters estimated from 
historical data and the additional source of the betting odds. We apply our analysis to a nine-year dataset of the most 
popular European leagues in order to predict match outcomes for their tenth seasons. In this paper, we provide numerical and graphical checks for our model.
}


\begin{document}

\maketitle

\section{Introduction}
\label{sec:intro}

In recent years, the challenge of modelling football outcomes has gained attention, in large part due to the potential for making substantial profits in betting markets. According to the current literature, this task may be achieved 
by adopting two different modelling strategies: the \textit{direct} models, for the number of goals scored by two competing teams; and the \textit{indirect} models, for estimating the 
probablility of the categorical outcome of a win, a draw, or a loss, which will hereafter be referred to as a \textit{three-way} process.   

The basic assumption of the direct models is that the number of goals scored by the two teams follow two Poisson distributions. Their dependence structure and the specification of their 
parameters are the other most relevant assumptions, according to the literature.
The scores' dependence issue is, in fact, the subject of much debate, and the discussion cannot yet be concluded. As one of the first contributors to the modelling of football scores, \cite{maher1982modelling} used two conditionally independent Poisson distributions, one for the goals scored by the home team, and another for the away team.
\cite{dixon1997modelling} expanded upon Maher's work and extended his model, introducing a parametric dependence between the scores. This also represents the justification for the bivariate Poisson 
model, introduced in \citet{karlis2003analysis} in a frequentist perspective, and in \cite{ntzoufras2011bayesian} under a Bayesian perspective. On the other hand, \cite{baio2010bayesian} assume the conditional independence within hierarchical Bayesian models, on the grounds that the correlation of the goals is already taken into account by the hierarchical 
structure. Similarly, \cite{groll2013spain} and \cite{groll2015prediction} show that, up to a certain amount, the scores' dependence on two competing teams may 
be explained by the inclusion of some specific teams' covariates in the linear predictors. However, \cite{dixon1998birth} note that modelling the dependence along a single match is possible: in such a case, a 
temporal structure in the 90 minutes is required.

The second common assumption is the inclusion in the 
models of some teams' effects to describe the attack and the defence strengths of the competing teams. Generally, they are 
used for modelling the scoring rate of a given team, and in much of the aforementioned literature they do not vary over time. Of course, 
this is a major limitation of these models. \cite{dixon1997modelling} tried to overcome this problem by downweighting the likelihood exponentially over time in order to reduce the impact of matches far from the current time of evaluation. However, over the last 10 years the advent of some dynamic models allowed these teams' effects to vary over the season, and to have a temporal structure. The independent Poisson model proposed by \cite{maher1982modelling} has been extended to a Bayesian dynamic independent model, where the evolution
structure is based on continuous time \citep{rue2000prediction}, or is specified for discrete times, such as a random walk for both the attack and defence parameters \citep{owen2011dynamic}. Instead the non-dynamic bivariate Poisson model is extended in \cite{koopman2015dynamic} and \cite{koopman2017forecasting}, and is expressed as a state space model where the teams' effects vary in function of a state vector.

For our purposes, the scores' dependence assumption may be relaxed, and in this paper we adopt a conditional independence 
assumption. From a purely conceptual point of view, we have several reasons for adopting two independent 
Poisson: (i) as discussed by \cite{baio2010bayesian}, assuming two conditionally independent Poisson hierarchical Bayesian 
models implicitely allows for correlation, since the observable variables are mixed at an upper level; (ii) as 
noted by \cite{mchale2011modelling}, there is empirical evidence that goals of two teams in seasonal leagues display 
only slightly positive correlation, or no correlation at all, whereas goals are negatively correlated for national teams; (iii) bivariate 
Poisson models \citep{karlis2003analysis}, which represent the most typical choice for modelling correlation, only allow for 
non-negative correlation. Moreover, the independence assumption allows for a simpler formulation for the likelihood 
function and simplifies the inclusion of the bookmakers' odds in our model. Concerning the dynamic assumption of the teams'-
specific effects, we use an autoregressive model by centring the effect of seasonal time $\tau$ at the lagged effect in $
\tau-1$, plus a fixed effect.

Whatever the choices for the two assumptions discussed above, the model proposed in this context was built with both a 
descriptive and a predictive goal, and its parameters' estimates/model probabilities were often used 
for building efficient betting strategies \citep{dixon1997modelling, londono2015sports}. In fact, the well 
known expression `beating the bookmakers' is often considered a mantra for whoever tries to predict football---or more 
generally, sports---results. As mentioned by \cite{dixon1997modelling}, to win money from the bookmakers
requires a determination of probabilities, which is sufficiently more accurate than those obtained from the odds. On the other hand, it is empirically known that betting odds are the most accurate source of information for forecasting sports 
performances \citep{vstrumbelj2014determining}. However, at least two issues deserve a deep analysis: how to determine probability forecasts from the raw 
betting odds, and how to use this source of information within a forecasting model (e.g., to predict the number of goals). 
Concerning the first point, it is well known that the betting odds do not correspond to probabilities; in fact, to make a profit, bookmakers set unfair odds, and they have a `take' of 5-10\%. In order to derive a set of coherent probabilities from these odds, many researchers have used the \textit{basic 
normalization} procedure, by normalising the inverse odds up to their sum. Alternatively, \cite{forrest2005odds} and \cite{forrest2002outcome} propose a 
regression model-based approach, modelling the betting probabilities through an historical set of betting odds and match outcomes.
But, \cite{vstrumbelj2014determining} shows that Shin's procedure \citep{shin1991optimal, shin1993measuring} gives the best results overall, being preferable both to the basic normalisation and regression approaches. 
Concerning the second issue, a small amount of literature focused on using the existing betting odds as \textit{part} of a 
statistical model for improving the predictive accuracy and the model fit. \cite{londono2015sports} used the 
betting odds for eliciting the hyperparameters of a Dirichlet distribution, and then updated them based on observations of the categorical 
three-way process. No researcher has tried to implement a similar strategy within the framework of the direct models.
\setlength{\parskip}{0pt}

In this paper we try to fill the gap, creating a bridge between the betting odds and betting probabilities on one hand and the statistical modelling of the scores.
Once we transform the inverse betting odds into probabilities, we will develop a procedure to (i) infer from these the implicit scoring intensities, according to the bookmakers, and
(ii) use these implicit intensities directly in the conditionally independent Poisson model for the scores,
within a Bayesian perspective. We are interested in both the estimation of the 
models parameters, and in the prediction of a new set of matches. Intuitively, the latter task is much more difficult than the 
former, since football is intrinsically noisy and hardly predictable. However, we believe that combining the 
betting odds with an historical set of data on match results may give predictions that are more accurate than those obtained from a single source of information.
\setlength{\parskip}{0pt}

In Section~\ref{sec:elicited} we introduce two methods, proposed by the current literature, for transforming the 
three-way betting odds favoured by bookmakers into probabilities. In Section~\ref{sec:model}, we introduce the full model, along with the implicit scoring rates. The results and predictive accuracy of 
the model on the top four European leagues---Bundesliga, Premier League, La Liga and Serie A---are presented in Section~\ref{sec:data}, and are summarised through posterior probabilities and 
graphical checks. Some profitable betting strategies are briefly presented in Section~\ref{sec:betting}. Section~\ref{sec:concl} concludes our analysis.

\section{Transforming the betting odds into probabilities}
\label{sec:elicited}

The connection between betting odds and probabilities has been broadly investigated over the last decades. Before proceeding, 
we will introduce the formal definition of odd and the related notation we are going to use throughout the rest of the paper. 
The odds of any given event are usually specified as the amount of money we would win if we bet one unit on that event. 
Thus, the odd 2.5 corresponds to 2.5 euro (or pounds) we would win betting 1 euro. The inverse odd---usually denoted 
as 1:2.5---corresponds to the unfair probability associated to that event. In fact, as is widely known, the 
betting odds do not correspond to probabilities: the sum of the inverse odds for a single match needs to be greater than one \citep{dixon1997modelling} in order to guarantee the bookmakers' profit. Here, $O_{m}= \{o_{Win}, o_{Draw}, o_{Loss} \} 
$, $
\Pi_{m}=(\pi_{Win}, \pi_{Draw}, \pi_{Loss} )$, and $\Delta_{m}=
\{ \mbox{`Win'}, 
\mbox{`Draw'}, \mbox{`Loss'} \}$ denote the vector of the inverse betting odds, the vector of the 
estimated betting probabilities, and the set of the three-way possible results for the $m$-th game, respectively. 

There is empirical evidence that the betting odds are the most accurate available source of 
probability forecasts for sports \citep{vstrumbelj2014determining}; in other words, forecasts based on odds-probabilities have been shown to be better, or at least as good as, statistical models, which use sport-specific predictors and/or expert tipsters.

However, some issues remain open. Among these is a strong debate over which method to 
use for inferring a set of probabilities from the raw betting odds. We can transform them into 
probabilities by using the two procedures proposed in the literature: the \textit{basic normalisation}---dividing the 
inverse odds by the booksum, i.e. the sum of the inverse betting odds, as broadly explained in \cite{vstrumbelj2014determining}---and \textit{Shin's procedure} 
described in \cite{shin1991optimal, shin1993measuring}. \cite{vstrumbelj2014determining}, \cite{cain2002one, 
cain2003favourite}, and \cite{smith2009bookmakers} show that Shin's probabilities improve over the basic normalisation: in \cite{vstrumbelj2014determining} this result has been achieved by the application of the Ranked Probability Score (RPS) \citep{epstein1969scoring}, which may be defined as a discrepancy measure between the probability of a three-way process outcome and the actual outcome. 

In this paper we will not focus focus on comparing these two procedures; rather, we are interested in using the probabilities derived from each for statistical and prediction purposes, as will become clearer in later sections.

\newpage

\begin{description}
\item[(A)]\textit{Basic normalisation}

\begin{equation}
\pi_{i}=\frac{o_{i}}{\beta}, \ i \in \Delta_{m},
\label{p:betting:dixon}
\end{equation}
where $\beta=\sum_{i}o_{i}$ is the so called booksum \citep{vstrumbelj2014determining}. The method has gained a great popularity due to its simplicity.

\item[(B)] \textit{Shin's procedure}

In the model proposed by \citet{shin1993measuring}, the bookmakers specify their odds in order to maximise their expected profit in a market with uninformed bettors and insider traders. The latter are those particular actors who, due to superior information, are assumed to \textit{already} know  the outcome of a given event---e.g. football match, horse race, etc.---before the event takes place. Their contribution in the global betting volume is quantified by the percentage $z$. \cite{jullien1994measuring} used Shin's model to explicitly work out the expression for the betting probabilities:

\begin{equation}
 \pi(z)_{i}= \frac{\sqrt{z^{2}+4(1-z)\frac{o_{i}^{2}}{\sum_{i} o_{i}}}-z}{2(1-z)} , \ i \in \Delta_{m},
 \label{p:betting:shin}
\end{equation}
so that $\sum_{i=1}^{3}\pi(z)_{i}=1$.
The current literature refers to these as Shin's probabilities.
The formula above is a function depending on the insider trading rate $z$, which \cite{jullien1994measuring} suggested should be estimated by nonlinear least squares as:
\[
 \mbox{Arg} \underset{z}{\min} \{\sum_{i=1}^{3}\pi(z)_{i}-1 \}.
 \]
 
The value here obtained may be defined as the minimum rate of insider traders that yields probabilities corresponding to the vector of inverse betting odds $O$.

\end{description}
\vspace{0.5cm}

\begin{figure}
\centering
\subfloat[Home win]{
\includegraphics[height=5cm, width=5cm]{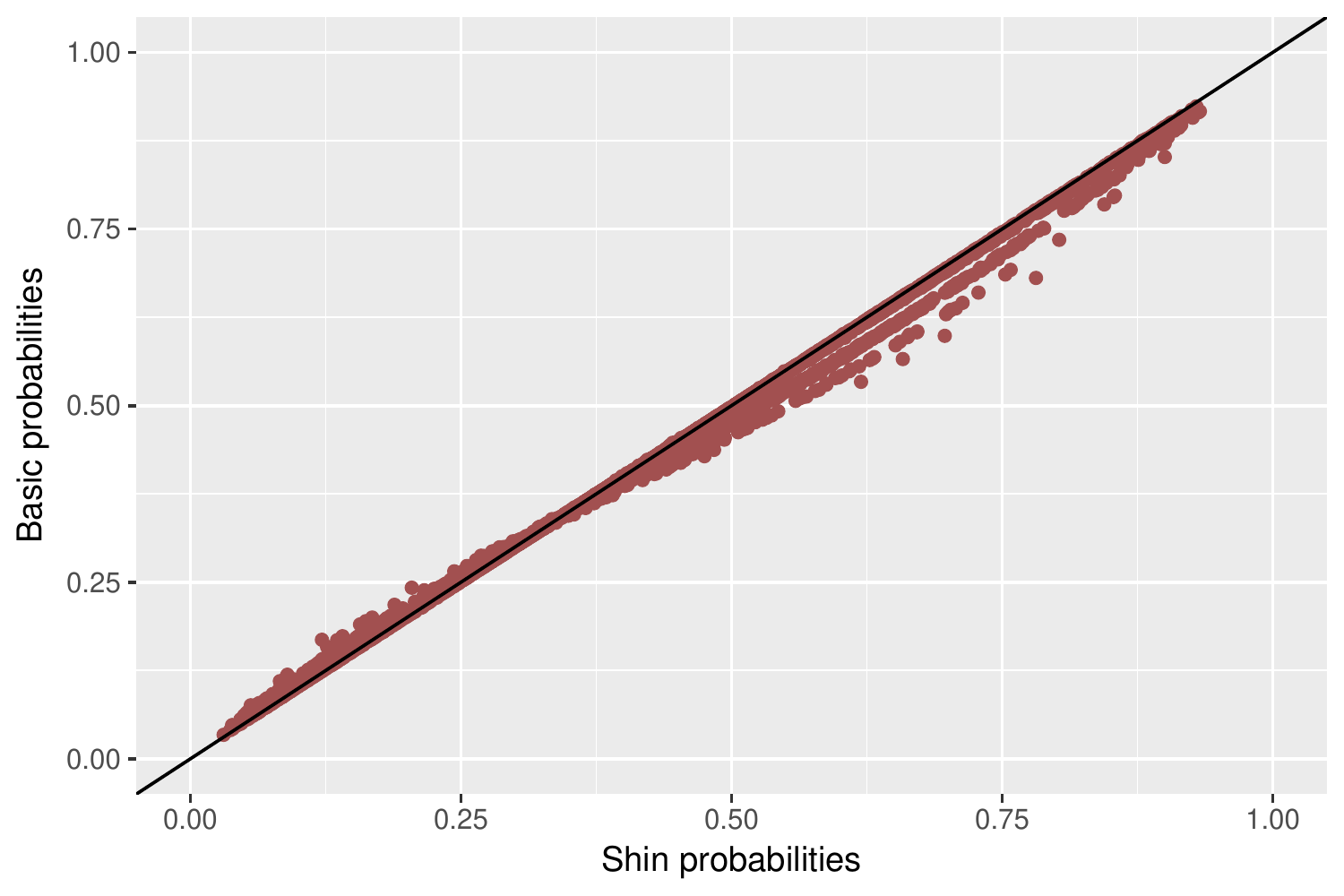}}~
\subfloat[Draw]{
\includegraphics[height=5cm, width=5cm]{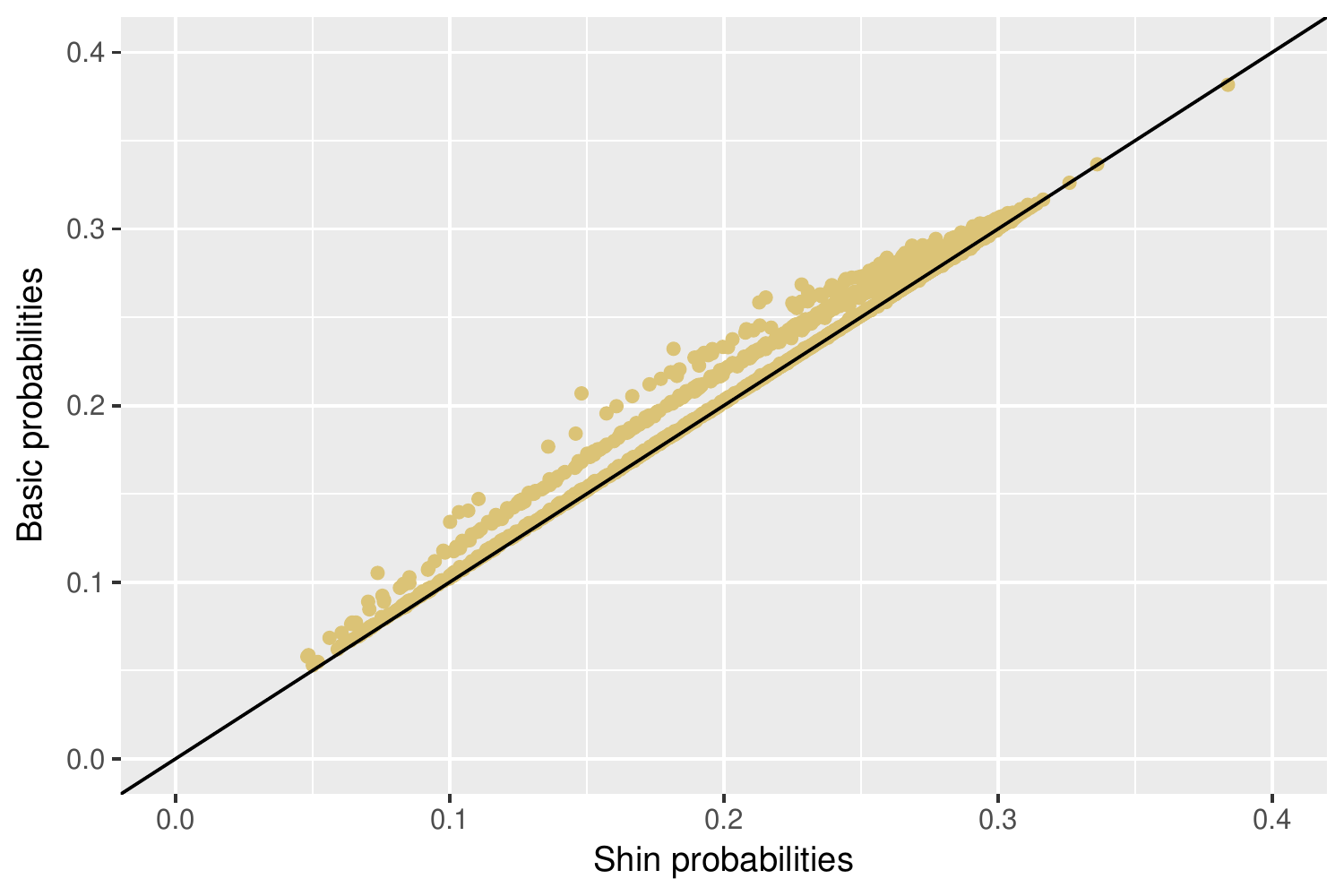}}~
\subfloat[Away win]{
\includegraphics[height=5cm, width=5cm]{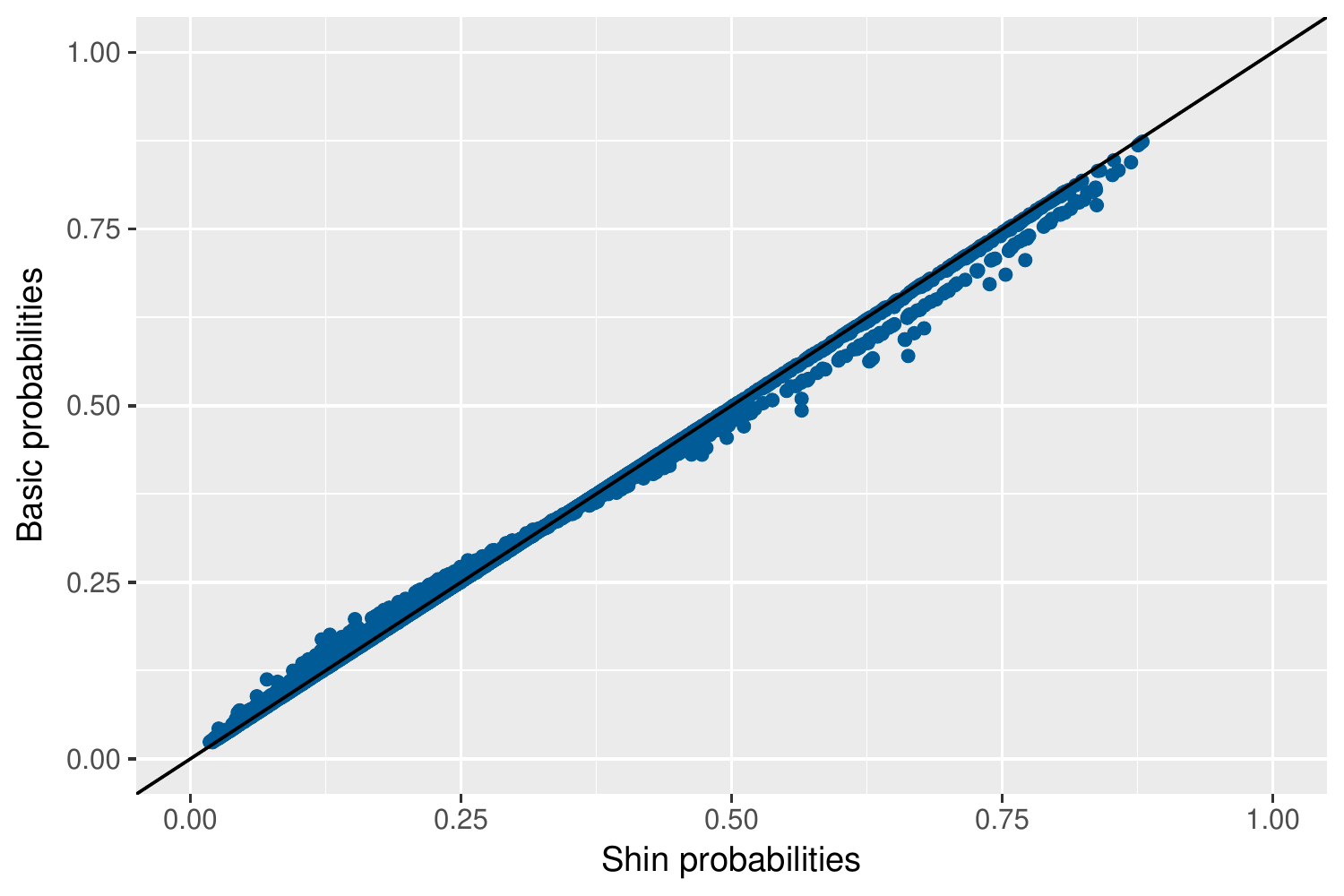}}
\caption{\label{fig01} Comparison between Shin probabilities ($x$-axis) and basic normalised probabilities ($y$-axis) for the Spanish La Liga championship (seasons from 2007/2008 to 2016/2017), according to seven different bookmakers.}
\end{figure}

Both of these methods yield probabilities, with the difference that Shin's procedure entails a function of the insider traders' 
rate which needs to be minimised for every match. Figure~\ref{fig01} displays the three-way betting probabilities 
obtained through the two procedures described above for La Liga (Spanish championship), from the season 2007-2008 to the 
season 2016-2017. As may be noted, the Draw probabilities obtained with the basic normalisation tend to be higher than 
those obtained with Shin's procedure. Conversely, as a home win and an away win tend to become more likely, Shin's procedure tends to favour them.

As is intuitive, a higher probability of a home win should somehow be associated with a greater number of goals scored by the home team, and the same for an away team.

\section{Model}
\label{sec:model}

\subsection{Model for the scores}

Here, $\bm{y}=(y_{m1}, y_{m2})$ denotes the vector of observed scores, where $y_{m1}$ and $y_{m2}$ are the number 
of goals scored by the home team and by the away team respectively in the $m$-th match of the dataset. 
 According to the motivations provided by \cite{baio2010bayesian}, in this paper we adopt a conditional 
 independence assumption between the scores. This choice allows for a simpler formulation for the likelihood function 
 and, later on, for the direct inclusion of the bookmakers' odds into the model through the Skellam distribution 
 \citep{karlis2009bayesian}.
 The model for the scores is then specified as:

\begin{align}
\begin{split}
y_{m1}| \theta_{m1} &\sim \mathsf{Poisson}(\theta_{m1})\\
y_{m2}| \theta_{m2} &\sim \mathsf{Poisson}(\theta_{m2}),\\
y_{m1}  \perp &y_{m2} | \theta_{m1}, \theta_{m2},
\end{split}
\label{y}
\end{align}
where $\bm{y}$ is modelled as \textit{conditionally} independent Poisson and the joint parameter $\bm{\theta}
=(\theta_{m1}, \theta_{m2}) $ represents the scoring intensities in the $m$-th game, for the home team and for the 
away team respectively. In what follows, we will refer to \eqref{y} as the \textit{basic} model, which is estimated 
using the past scores. The main novelty of this paper consists of enriching this specification by including the extra 
information which stems from the bookmakers' betting odds. Thus, for each pair of match $m$ and bookmaker $s, \ s=1,...,S
$ the betting probabilities $\pi^{s}_{i,m},\ i \in \Delta_{m}$, derived with one of the methods in Section~\ref{sec:elicited}, may be used to find out the values $\hat{\bm{\theta}}^{s}=(\hat{\theta}^{s}
_{m1}, \hat{\theta}^{s}_{m2})$, which solve the following nonlinear system of equations:
\vspace{0.3cm}

\begin{align}
\begin{split}
\pi^{s}_{Win, m}+\pi^{s}_{Draw, m}=& P( y_{m1} \ge y_{m2}| \theta^{s}_{m1}, \theta^{s}_{m2} ) \\
\pi^{s}_{Loss, m}=&P( y_{m1}< y_{m2}| \theta^{s}_{m1}, \theta^{s}_{m2} ).\\
\end{split}
\label{system}
\end{align}
\vspace{0.3cm}
The existence of these values is guaranteed by the fact that, under~\eqref{y}, $y_{m1}-y_{m2} \sim PD(\theta_{m1}, 
\theta_{m2})$, where $PD$ denotes the Poisson-Difference 
distribution, also known as Skellam distribution, with parameters $\theta_{m1}, \theta_{m2}$ and mean $\theta_{m1}-\theta_{m2}$. In such a way, we obtain for each pair $(m,s)$  the \textit{implicit} scoring rates  $\hat{\theta}^{s}
_{m1}, \hat{\theta}^{s}_{m2}$, somehow  inferring the scoring intensities implicit in the three-way bookmakers' odds. Now, we consider our augmented dataset by including as auxiliary data the observed $\hat{\theta}^{s}
_{m1}, \hat{\theta}^{s}_{m2}$. For every $m$, our new data vector is represented by:

$$ (\bm{y}, \bm{\hat{\theta}}^{s})=(y_{m1}, y_{m2}, \hat{\theta}^{s}
_{m1}, \hat{\theta}^{s}_{m2},\ s =1,...,S ).$$ 
Now, from Equation~\eqref{y} we move to the following specification:

\begin{align}
\begin{split}
y_{m1}|\theta_{m1}, \lambda_{m1} &\sim  \mathsf{Poisson}(p_{m1}\theta_{m1}+(1-p_{m1})\lambda_{m1})\\
y_{m2}|\theta_{m2}, \lambda_{m2} &\sim  \mathsf{Poisson}(p_{m2}\theta_{m2}+(1-p_{m2})\lambda_{m2}),
\end{split}
\label{y:mixture}
\end{align}
where $\lambda_{m1}, \ \lambda_{m2}$ are bookmakers' parameters introduced for modelling the additional data $\hat{\theta}^{s}
_{m1}, \hat{\theta}^{s}_{m2},\ s =1,...,S$, as explained in the next section. Parameters $p_{m1}, p_{m2}$ are assigned a non-informative prior distribution, with hyper-parameters $a$ and $b$, e.g. $p_{m\cdot} \sim \mathsf{Beta}(a,b)$.

\subsection{Model for the rates}

Equation~\eqref{y:mixture} introduced a convex combination for the Poisson parameters, accounting for both the scoring rates $\theta_{\cdot 1}, \theta_{\cdot 2}$ and the bookmakers' parameters $\lambda_{\cdot 1}, \lambda_{\cdot 2}$. Denoting with $T$ the number of teams, the common specification for the scoring intensities is a log-linear model in which for each $t, \ t=1,...,T$:

\begin{align}
\begin{split}
\log(\theta_{m1})&=\mu+att_{t[m]1}+def_{t[m]2}\\
\log(\theta_{m2})&=att_{t[m]2}+def_{t[m]1} 
\end{split}
\label{theta}
\end{align}
with the nested index $t[m]$ denoting the team $t$ in the $m$-th game. The parameter $\mu$ represents the well-known 
football advantage of playing at home, and is assumed to be constant for all the teams over time, as in the current 
literature. The attack and defence strengths of the competing teams are modelled by the parameters $att$ and $def$ 
respectively. \cite{baio2010bayesian} and \cite{dixon1997modelling} assume that these team-specific 
effects do not vary over the time, and this represents a major limitation in their models. In fact, \cite{dixon1998birth} 
show that the attack and defence effects are not static and and may even vary during a single match; thus, a static 
assumption is often not reliable for making predictions and represents a crude approximation of the reality. 
\cite{rue2000prediction} propose a generalised linear Bayesian model in which the team-effects at match time $\tau$ 
are drawn from a Normal distribution centred at the team-effects at match time $\tau-1$, and with a variance term 
depending on the time difference. We make a seasonal assumption 
considering the effects for the season $\tau$ following a Normal distribution centred at the previous seasonal effect 
plus a fixed component. For each $t=1,\ldots,T, \ \tau=2,\ldots,\mathcal{T}$:

\begin{align}
\begin{split}
 att_{t,\tau} &\sim \mathsf{N}(\mu_{att}+att_{t, \tau-1}, \sigma^{2}_{att}) \\
 def_{t, \tau} &\sim \mathsf{N}(\mu_{def}+def_{t, \tau-1}, \sigma^{2}_{def}), 
 \end{split}
 \label{att:def}
 \end{align}
while, for the first season, we assume: 

\begin{align}
\begin{split}
 att_{t,1} &\sim \mathsf{N}(\mu_{att}, \sigma^{2}_{att}) \\
 def_{t, 1} &\sim \mathsf{N}(\mu_{def}, \sigma^{2}_{def}). 
 \end{split}
 \label{att:def:1}
 \end{align}
As outlined in the literature, we need to impose a `zero-sum' identifiability constraint within each season to these random effects:

$$ \sum_{t=1}^{T} att_{t, \tau}=0, \ \ \  \ \sum_{t=1}^{T}def_{t, \tau}=0, \ \  t=1,\ldots, T, \ \tau=1,\ldots \mathcal{T}, $$
whereas $\mu$ and the hyperparameters of our model are assigned weakly informative priors:

\vspace{-0.5cm}

\begin{align*}
\mu, \mu_{att}, \mu_{def} \sim & \mathsf{N}(0,10)\\
\sigma_{att}, \sigma_{def} \sim & \mathsf{Cauchy}^{+}(0,2.5),\\
 \end{align*}
where $\mathsf{Cauchy^+}$ denotes the half-Cauchy distribution, centred in 0 and with scale 2.5.\footnote{On the
choice of the half-Cauchy distribution for scale parameters, see
\citet{gelman2006prior}.} The team-specific effects modelled through Equations~\eqref{att:def} and~\eqref{att:def:1} are estimated from the past scores in the dataset. As expressed in \eqref{y:mixture}, we add a level to the hierarchy, by including the implicit scoring rates as a separate data model. Given, then, a further level which consists of $S$ bookmakers, it is natural to consider $ \lambda_{m1}, \lambda_{m2}$ as the model parameters for the observed $\hat{\theta}^{s}_{m1}, \hat{\theta}^{s}_{m2}$. More precisely, these parameters represent the means
of two truncated Normal distributions for the further implicit scoring rates model:

\begin{align}
\begin{split}
\hat{\theta}^{1}_{m1},...,\hat{\theta}^{S}_{m1} & \sim \mbox{trunc}\mathsf{ N}( \lambda_{m1}, \tau^{2}_{1}, 0, \infty)\\
\hat{\theta}^{1}_{m2},...,\hat{\theta}^{S}_{m2} & \sim \mbox{trunc}\mathsf{ N}( \lambda_{m2}, \tau^{2}_{2}, 0, \infty),
\end{split}
\label{theta_bm}
\end{align}
where $ \mbox{trunc}\mathsf{ N}(\mu, \sigma^{2}, a,b) $ is the 
common notation for the density of a truncated Normal with 
parameters $\mu \in \mathbb{R},  \sigma^{2} \in \mathbb{R}^{+}$ 
and defined in the interval $[a,b]$. $\lambda_{m1}, \lambda_{m2}
$ are in turn assigned two truncated Normal distributions:

\begin{align}
\begin{split}
\lambda_{m1}& \sim \mbox{trunc} \mathsf{ N}( \alpha_{1}, 10, 0, \infty)\\
\lambda_{m2} & \sim \mbox{trunc}\mathsf{  N}( \alpha_{2} , 10, 0, \infty),
\end{split}
\label{lambda_bm}
\end{align}
with hyperparameters $\alpha_{1}, \alpha_{2}$.

\section[Applications and results]{Applications and results: top four European leagues }
\label{sec:data}

\subsection{Data}

We collected the exact scores for the top four European professional leagues---Italian Serie A, English Premier League, 
German Bundesliga, and Spanish La Liga---from season 2007/2008 to 2016/2017. Moreover, we also collected all the three-way odds for 
the following bookmakers: Bet365, Bet\&Win, Interwetten, Ladbrokes, Sportingbet, VC Bet, William Hill. All these data have 
been downloaded from the public available page \url{http://www.football-data.co.uk/}. We are interested in both (a) 
posterior predictive checks in terms of replicated data under our models, and (b) out-of-sample predictions for a new dataset. 
According to point (b), which appears to be more appealing for fans, bettors and statisticians,  let $\mathcal{T}_{r}$  denote 
the \textit{training set}, and $\mathcal{T}_{s}$ the \textit{test set}. Our training set contains the results of nine seasons for each 
professional league, and our test set contains the results of the tenth season.

The model coding has been implemented in {\ttfamily WinBUGS}
\citep{spiegelhalter2003winbugs} and in {\ttfamily Stan} \citep{rstan}. We ran our MCMC simulation for $H=5000$ 
iterations, with a burn-in period of $1000$, and we monitored the convergence using the usual MCMC diagnostic \citep{gelman2014bayesian}.

\subsection{Parameter estimates}

\begin{sidewaysfigure}
\centering
\makebox{\includegraphics[scale=1.3]{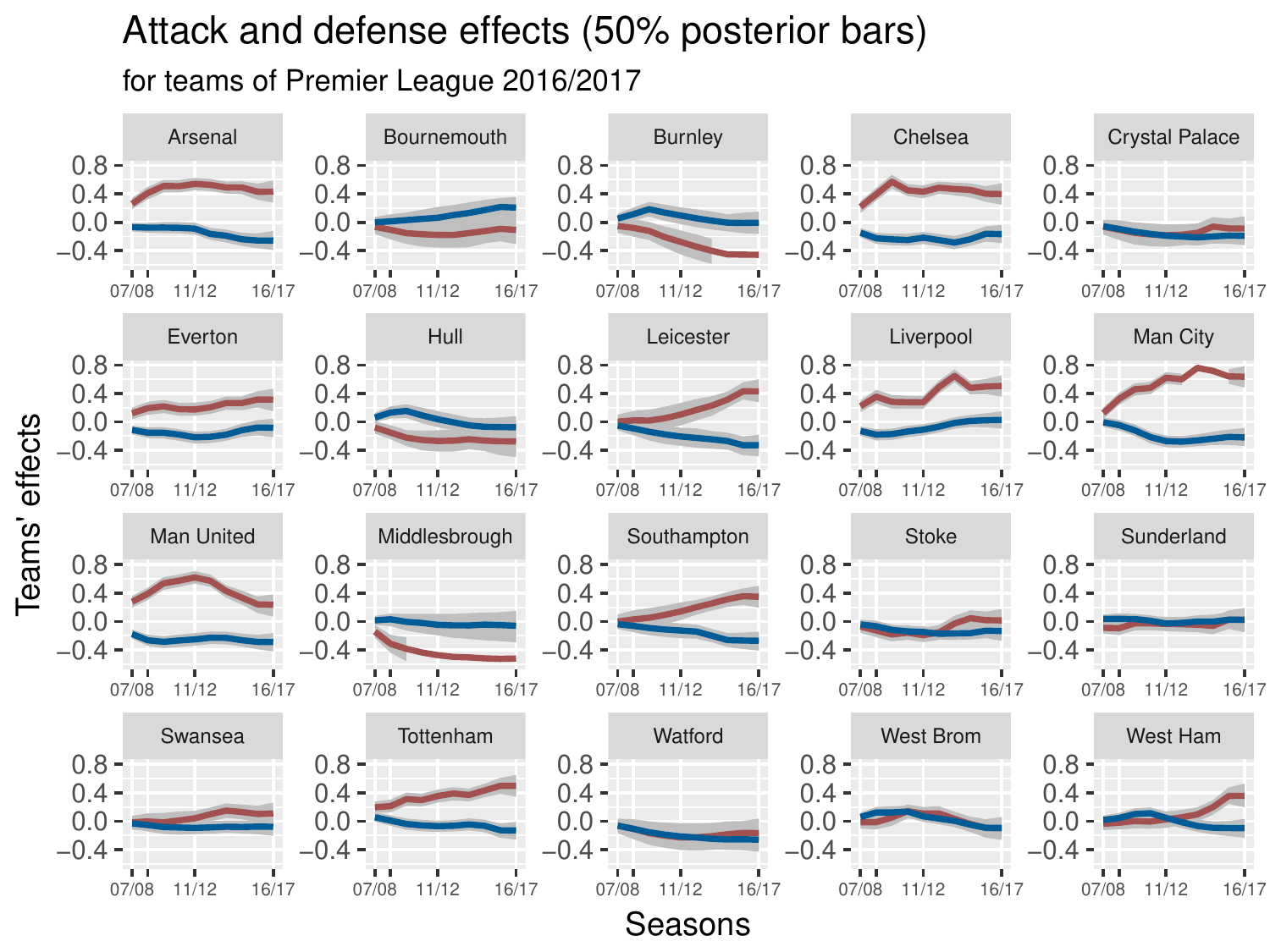}}
\caption{\label{fig02b} Posterior 50\% confidence bars  for the attack (red) and the defence (blue) effects along the 10 seasons for the teams belonging to the Premier League 2016/2017. Wider posterior bars are associated with teams reporting fewer observations.}
\end{sidewaysfigure}

As broadly explained in Section~\ref{sec:model}, the model in~\eqref{y:mixture} combines historical information about the scores and betting information about the odds. We acknowledge that the scoring rate is a convex combination that \textit{borrows strengths} from both the sources of information.
%\textcolor{red}{Grafico parametro p o tabella o qualcosa del genere?}
Figure~\ref{fig02b} displays the posterior estimates for the attack and the defence parameters associated 
with the teams belonging to the English Premier League during the test set season 2016-2017. The larger is the team-
attack parameter, and the greater is the attacking quality for that team; conversely, the lower is the team-
defence parameter, and the better is the defence power for that team. As a general comment, after reminding the 
reader that these quantities are estimated using only the historical results,  the pattern seems to reflect the actual 
strength of the teams across the seasons. For example Chelsea and Manchester City register the highest effects for the attack and the 
lowest for the defence across the nine seasons considered: consequently, the out-of-sample estimates for the tenth season 
mirror previous performance. Conversely, weaker teams are associated with an inverse pattern: see for instance Hull City, Middlesbrough, and Sunderland, all relegated at the end of the season. It is worth noting 
that some wide posterior bars are associated to those teams with fewer seasonal observations: in fact, for simplicity, we 
do not account for a relegation system, and some teams have been observed less during the seasons considered.

\begin{figure}
\centering
\subfloat[$p_{\cdot 1}$]
{\includegraphics[scale=0.45]{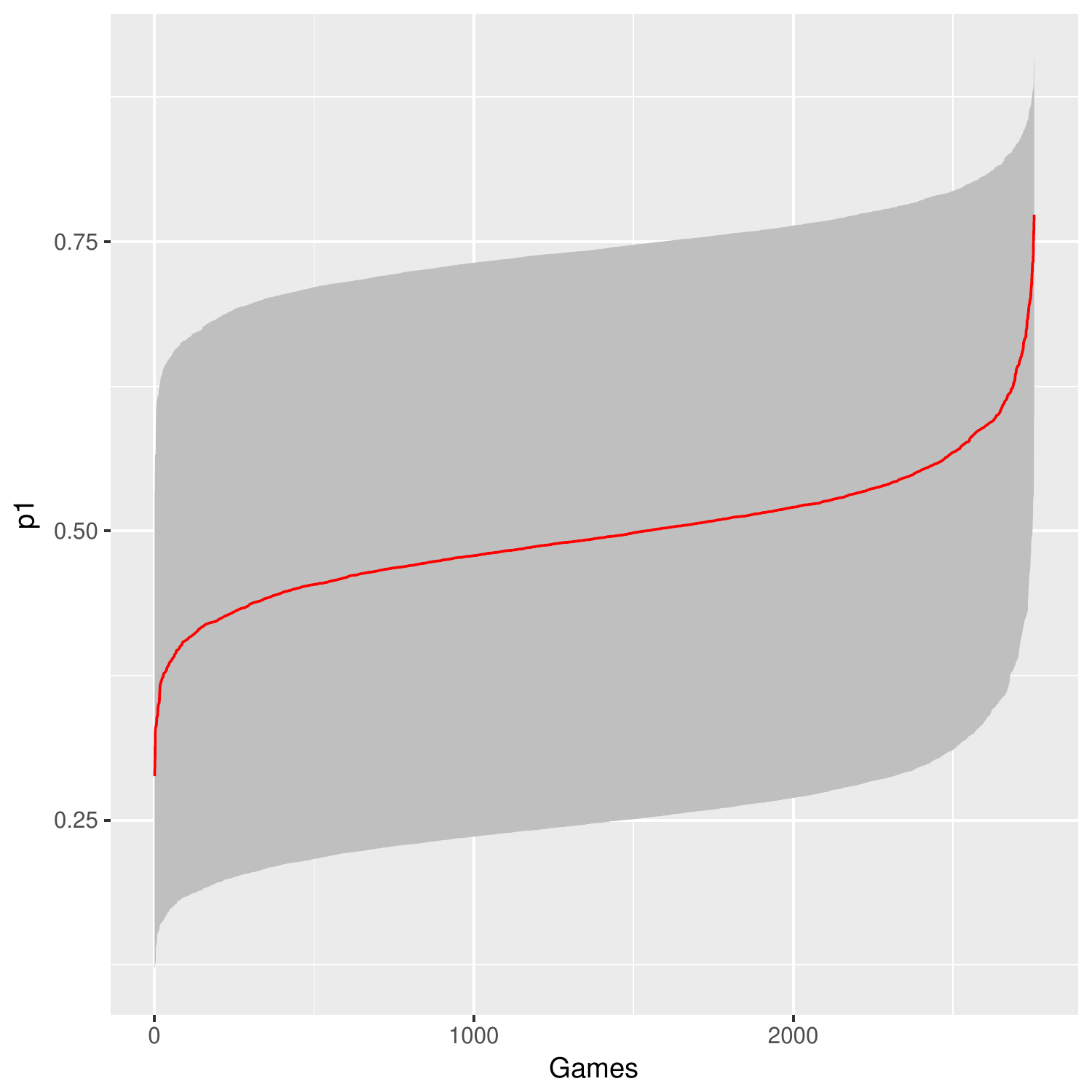}}~
\subfloat[$p_{\cdot 2}$]
{\includegraphics[scale=0.45]{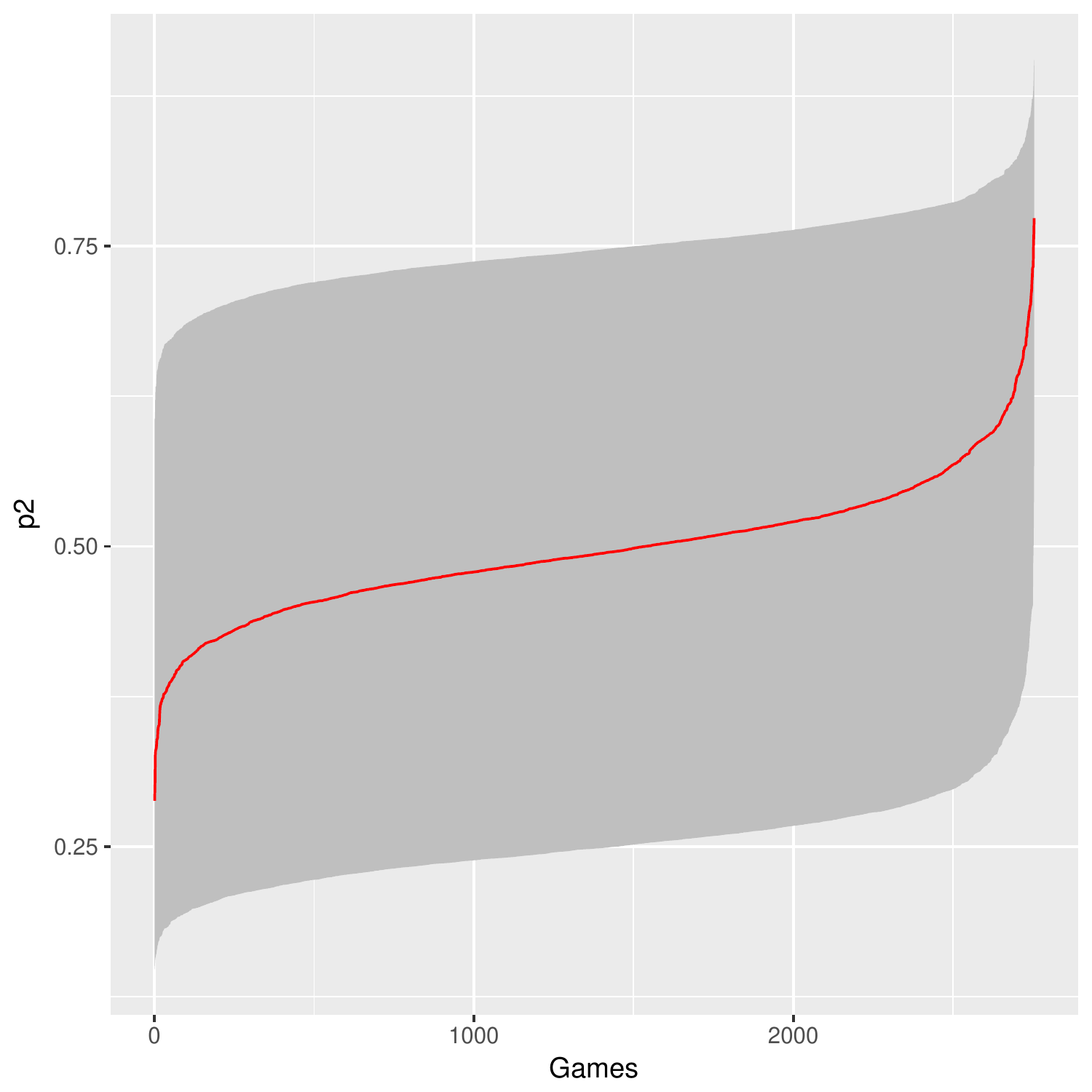}}
\caption{\label{fig03p} Ordered posterior 50\% confidence bars for parameters $p_{\cdot 1}, \ p_{\cdot 2}$ for German Bundesliga (from 2007-2008 to 2015-2016), 2754 matches.}
\end{figure}

Figure~\ref{fig03p} displays the ordered 50\% confidence bars for the marginal posteriors of the probabilities parameter $p_{m1}, p_{m2}, m=1,\ldots,M$, which appear in~\eqref{y:mixture}, computed for the German Bundesliga. Despite the high variability, these plots suggest that the amount of information that stems from the bookmakers is comparable with that arising from historical information. Then, the convex combination in~\eqref{y:mixture} seems to be an adequate option for our purposes.

\subsection{Model fit}

As broadly explained in \cite{gelman2014bayesian}, once we obtain some estimates from a Bayesian model we should assess the fit of this 
model to the data at hand and the plausibility of such model, given the purposes for which it was built. The 
principal tool designed for achieving this task is \textit{posterior predictive checking}. This post-model 
procedure consists of verifying whether some additional replicated data under our model are similar to the observed 
data. Thus, we draw simulated values $y^{rep}$ from the joint predictive distribution of replicated data:

\begin{equation*}
p(y^{rep}|y) = \int_{\Theta} p({y}^{rep}, \theta|y) d\theta= \int_{\Theta} p(\theta|y)p({y}^{rep}|\theta) d\theta.
\label{posterior_predicitve_1}
\end{equation*}
It is worth noting that the symbol $y^{rep}$ used here is different from the symbol $\tilde{y}$ used in the next section. The former is just a replication of $y$, the latter is any future observable value. 

Then, we define a test statistic $T(y)$ for assessing the discrepancy between the model and the data. A lack of fit of the model with respect to the posterior predictive distribution may be measured by tail-area posterior probabilities, or Bayesian $p$-values

\begin{equation}
p_{B}= P(T(y^{rep})>T(y)|y).
\label{eq:bayesian_p}
\end{equation}
As a practical utility, we usually do not compute the integral in~\eqref{posterior_predicitve_1}, but compute the posterior 
predictive distribution through simulation. If we denote with $\theta^{(s)}, \ s=1,...,S$ the $s$-th MCMC draw  from the 
posterior distribution of $\theta$, we just draw $y^{rep}$ from the predictive distribution $p(y^{rep}| \theta^{(s)})$. 
Hence, an estimate for the Bayesian $p$-value is given by the proportion of the $S$ simulations for which the quantity 
$T(y^{rep \ (s)})$ exceeds the observed quantity $T(y)$. From an interpretative point of view, an extreme $p$-value---too 
close to 0 or 1---suggests a lack of fit of the model compared to the observed data. 

\begin{figure}[ht]
\centering
\subfloat[Bundesliga]
{\includegraphics[scale=0.5]{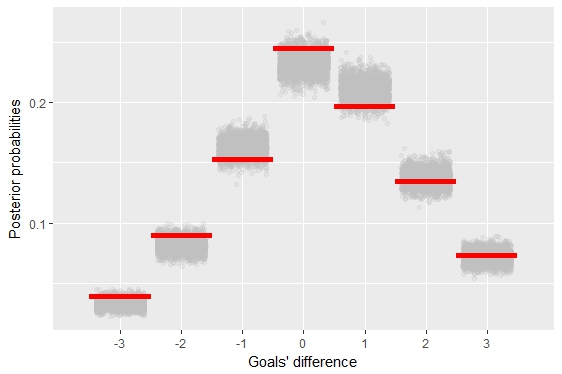}}~
\subfloat[La Liga]
{\includegraphics[scale=0.5]{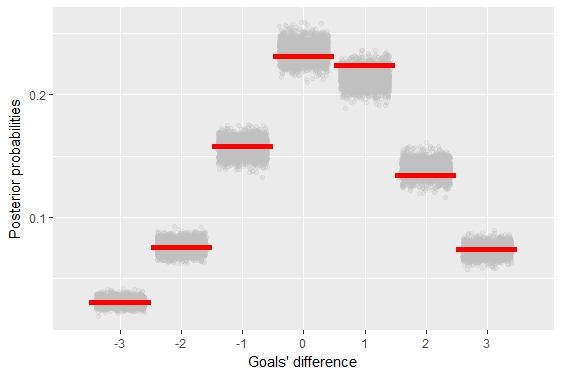}}\\
\subfloat[Premier League]
{\includegraphics[scale=0.5]{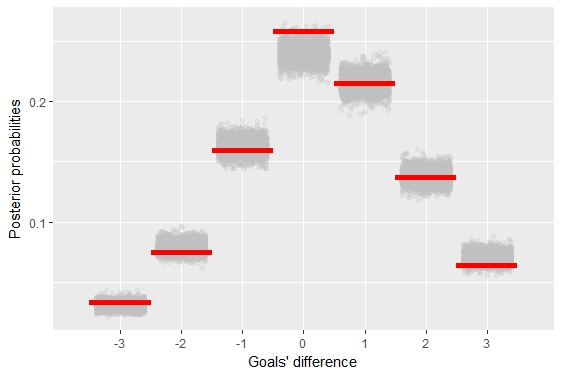}}~
\subfloat[Serie A]
{\includegraphics[scale=0.5]{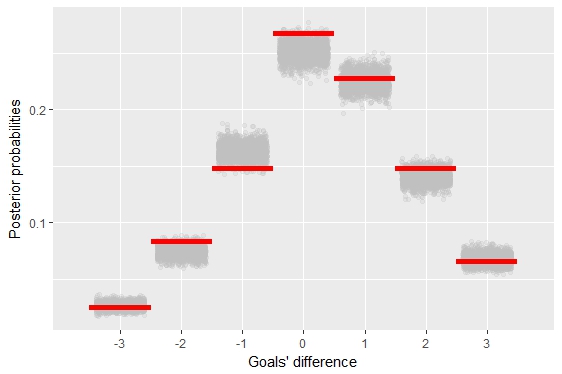}}\\
\caption{\label{fig03dens} PP check for the goals' difference $y_{1}-y_{2}$ against the replicated goals' difference $y^{rep}_{1}-y^{rep}_{2}$ for the top four European leagues . For each league, the graphical posterior predictive checks show that the model fits the data well.}
\end{figure}

Rather than comparing the posterior distribution of some statistics with their observed values 
\citep{gelman2014bayesian}, we propose a slightly different approach, allowing for a broader comparison of the replicated 
data under the model. Figure~\ref{fig03dens} displays the replicated distributions $y^{rep}_{1}-y^{rep}_{2}$ (grey 
areas) and the observed goals' difference (red horizontal line) from the top four European 
leagues. From this plots the fit of the model seems good: in other words, the replicated data under the model are plausible 
and close to the data at hand. As it may be noted, the variability of the replicated goals' difference amounting to 
-1, 0, 1 is greater than the variability for a goals' difference of -3 or 3. Moreover, the observed goals' 
differences always fall within the replicated distributions. In correspondence of a draw---goal difference of 0---the 
observed goals' differences register an high posterior probability if compared with the corresponding replicated distribution. 

\subsection{Prediction and posterior probabilities}

\begin{figure}
\centering
%\makebox{\includegraphics[scale=0.6]{RealBarca.pdf}~
%\includegraphics[scale=0.6]{SampJuve.pdf}
%}\\
\makebox{
\includegraphics[scale=0.45]{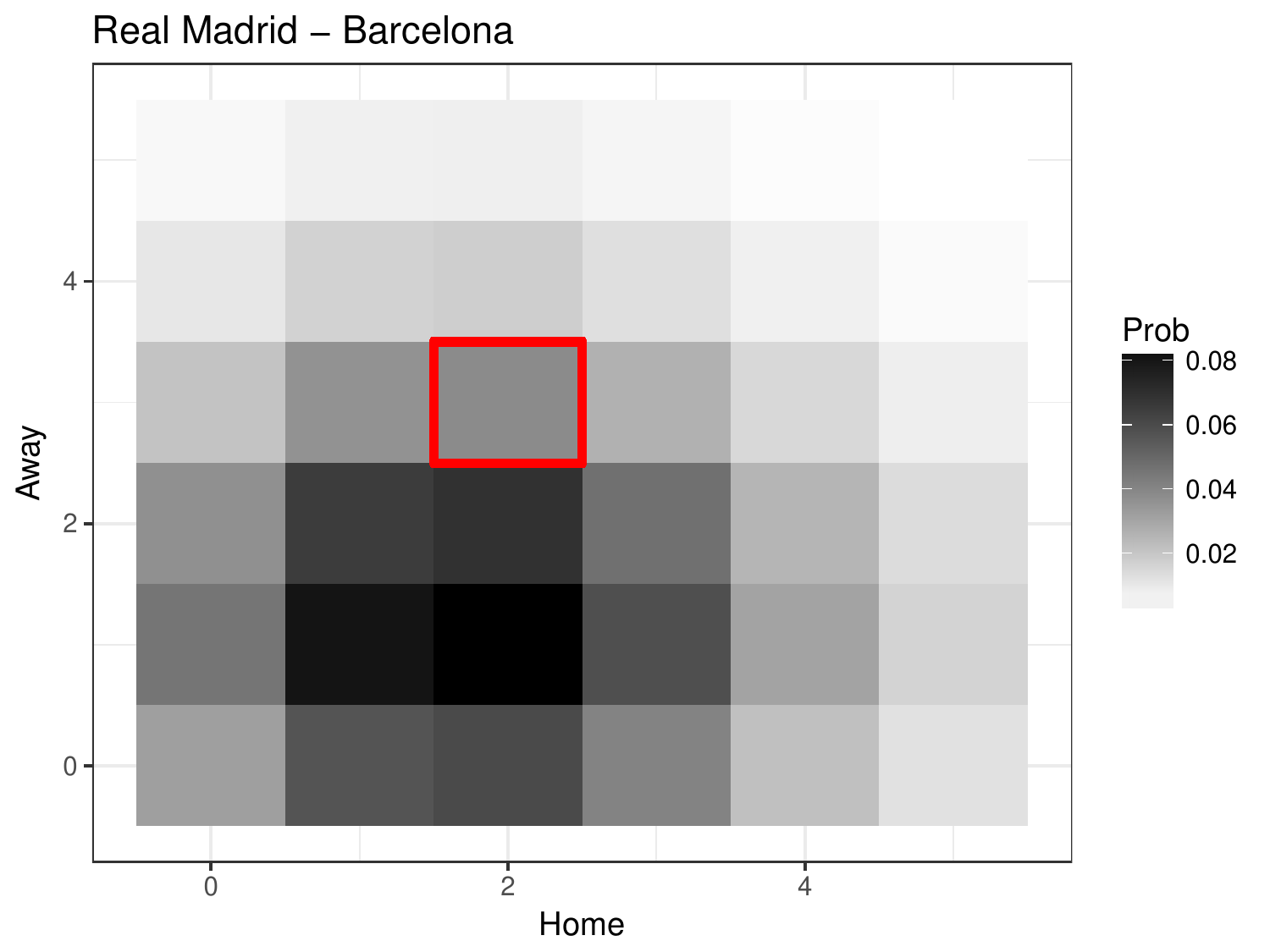}~
\includegraphics[scale=0.45]{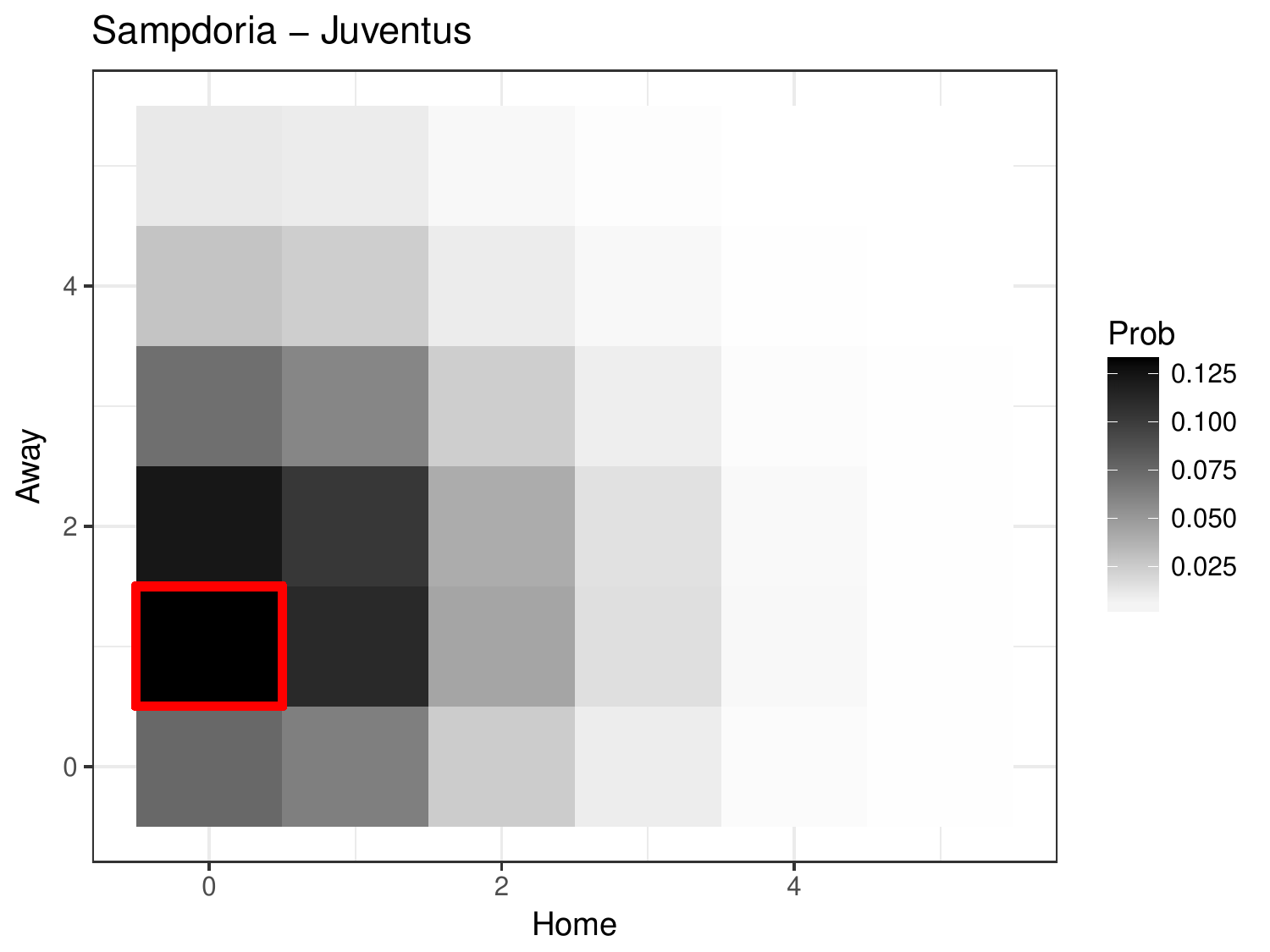}}
\caption{\label{fig03} Posterior predictive distribution of the possible results for the match Real Madrid-Barcelona, Spanish La Liga 2016/2017, and Sampdoria-Juventus, Italian Serie A 2016-2017. Both the plots report the posterior uncertainty related to the exact predicted outcome. Darker regions are associated with higher posterior probabilities and red square corresponds with the observed result.}
\end{figure}

The main appeal of a statistical model relies on its predictive accuracy. As usual in a Bayesian framework, the prediction for a new dataset may be performed directly via the 
posterior predictive distribution for our unknown set of observable values. Following the same notation of 
\cite{gelman2014bayesian}, let us denote with $\tilde{y}$ a generic unknown observable. Its distribution is then conditional on the observed $y$,

\begin{equation*}
p(\tilde{y}|y) = \int_{\Theta} p(\tilde{y}, \theta|y) d\theta= \int_{\Theta} p(\theta|y)p(\tilde{y}|\theta) d\theta,
\label{posterior_predicitve_2}
\end{equation*}
where the conditional independence of $y$ and $\tilde{y}$ given $\theta$ is assumed. Figure~\ref{fig03} displays the posterior predictive 
distributions for Real Madrid-Barcelona, Spanish La Liga 2016/2017, and for Sampdoria-Juventus, Italian Serie A. 
The red square indicates the observed result, (2,3) for the first match and (0,1) for the second match respectively.  
Darker regions are associated with higher posterior probabilities. According to the model, the most likely result for the first game is (2,1), with an associated 
posterior probability slightly greater than 0.08, whereas the most likely result coincide with the actual result (0,1) for the second game. 

These plots are not actually suggesting the most likely result: would it be smart to bet on an event with an associated probability about 0.09? Maybe, not. Rather, these plots provide 
a picture that acknowledges the large uncertainty of the prediction. We are not really interested in a model 
that often indicates a rare result that has been observed as the most likely outcome; we suspect, in fact, that a model which would 
favour the outcome (2,3) as most (or quite) likely, is probably not a good model. Rather, being aware of the unpredictable nature 
of football, we would like to grasp the posterior uncertainty of a match outcome in such a way that the actual result is not extreme in the predictive distribution.

Table~\ref{tab01} and Table~\ref{tab02} report the estimated posterior probabilities for each team being the first, 
the second, and the third; the first relegated, the second relegated, and the third relegated for each of the top four 
leagues, together with the observed rank and the achieved points, respectively. At the beginning of the 2016-2017 season, 
Bayern Munich had an estimated probability 0.8168 of winning the German 
league, which it actually did; in Italy, Juventus had an high probability of being the first (0.592) as well. Conversely, 
Chelsea had  a low associated probability to win the English Premier League at the beginning of the season, and this is mainly due to the bad 
results obtained by Chelsea in the previous season. Of course, the model does not account for the players'/managers' transfer market occurring in the summer period. In July 2016, 
Chelsea hired Antonio Conte, one of the best European managers, who won the English Premier League on his first attempt. For the relegated teams, it is worth noting that Pescara has high estimated probability to be the worst 
team of the Italian league (0.46). Globally, the model appears able to identify the teams with an associated high relegation's posterior probability.

\begin{table}
\caption{\label{tab01} Estimated posterior probabilities for each team being the first, the second, and the third in the Bundesliga,  Premier League, La Liga and Serie A 2016-2017, together with the observed rank and the number of points achieved.}
\centering
\bgroup
\def\arraystretch{0.7}
\begin{tabular}{lcccccc}
  \hline
&Team & P(1st) & P(2nd) & P(3rd) & Actual rank & Points \\ 
  \hline
  \multirow{3}{*}{\includegraphics[scale=0.05]{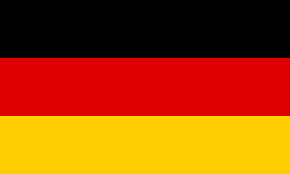}}&
   Bayern Munich & 0.8168 & 0.1508 & 0.0248 & 1 & 82 \\ 
   &RB Leipzig & 0.008 & 0.0284 & 0.0608 & 2 & 67 \\ 
   &Dortmund & 0.1332 & 0.4712 & 0.1856 & 3 & 64 \\ 
   \hline
   \multirow{3}{*}{\includegraphics[scale=0.05]{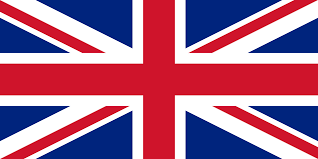}}&
   Chelsea & 0.1396 & 0.1592 & 0.1584 & 1 & 93 \\ 
   &Tottenham & 0.1096 & 0.132 & 0.1424 & 2 & 86 \\ 
   &Man City & 0.3904 & 0.2004 & 0.1388 & 3 & 78 \\ 
   \hline
   \multirow{3}{*}{\includegraphics[scale=0.05]{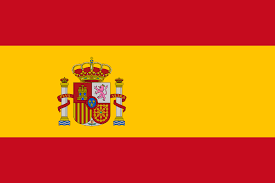}}&
 Real Madrid & 0.3868 & 0.4844 & 0.1076 & 1 & 93 \\ 
  & Barcelona & 0.5652 & 0.3536 & 0.0728 & 2 & 90 \\ 
   &Ath Madrid & 0.046 & 0.1348 & 0.5556 & 3 & 78 \\ 
   \hline
   \multirow{3}{*}{\includegraphics[scale=0.05]{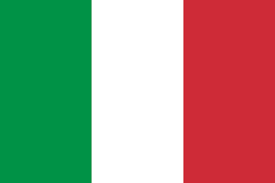}}&
      Juventus & 0.592 & 0.2335 & 0.107 & 1 & 91 \\ 
   &Roma & 0.1535 & 0.263 & 0.2595 & 2 & 87 \\ 
   &Napoli & 0.206 & 0.2965 & 0.213 & 3 & 86 \\ 
   \hline
\end{tabular}
\egroup
\end{table}

\begin{table}
\caption{\label{tab02} Estimated posterior probabilities for each team being the first, the second, and the third relegated team in the Bundesliga, Premier League, La Liga, and Serie A 2016-2017, together with the observed rank and the number of points achieved.}
\centering
\bgroup
\def\arraystretch{0.7}
\begin{tabular}{lcccccc}
  \hline
 &Team & P(1st rel) & P(2nd rel) & P(3d rel) & Actual rank & Points \\ 
  \hline
  \multirow{3}{*}{\includegraphics[scale=0.05]{deutche.png}}& Wolfsburg & 0.0212 & 0.0236 & 0.0064 & 16 & 37 \\ 
   &Ingolstadt & 0.0952 & 0.0904 & 0.0912 & 17 & 32 \\ 
   &Darmstadt & 0.1192 & 0.1552 & 0.2528 & 18 & 25 \\ 
   \hline
   \multirow{3}{*}{\includegraphics[scale=0.05]{inglese.png}}& Hull & 0.1384 & 0.1512 & 0.1428 & 18 & 34 \\ 
  & Middlesbrough & 0.118 & 0.1448 & 0.1812 & 19 & 28 \\ 
  & Sunderland & 0.1272 & 0.1228 & 0.1144 & 20 & 24 \\ 
  \hline
   \multirow{3}{*}{\includegraphics[scale=0.05]{spagnola.png}}&
  Sp Gijon & 0.1132 & 0.1112 & 0.1016 & 18 & 31 \\ 
   &Osasuna & 0.1464 & 0.174 & 0.228 & 19 & 22 \\
    &Granada & 0.138 & 0.1748 & 0.2476 & 20 & 20 \\ 
    \hline
    \multirow{3}{*}{\includegraphics[scale=0.05]{italiana.png}}& 
    Empoli & 0.0795 & 0.066 & 0.0415 & 18 & 32 \\ 
   &Palermo & 0.132 & 0.1765 & 0.1205 & 19 & 26 \\ 
   &Pescara & 0.1215 & 0.178 & 0.46 & 20 & 18 \\ 
   
   \hline
\end{tabular}
\egroup
\end{table}

\begin{figure}
\centering
\subfloat[Bundesliga]
{\includegraphics[scale=0.5]{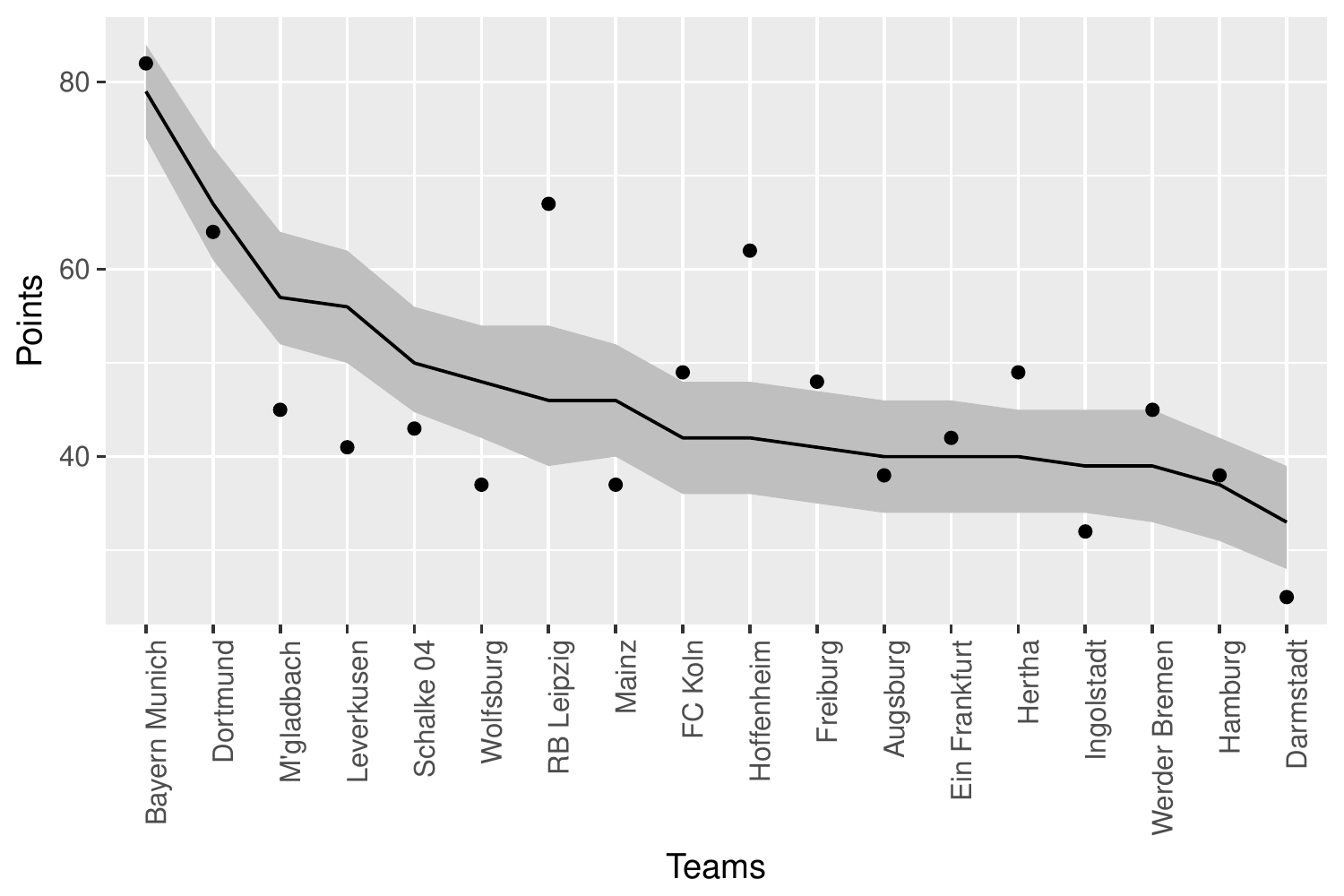}}~
\subfloat[Premier League]
{\includegraphics[scale=0.5]{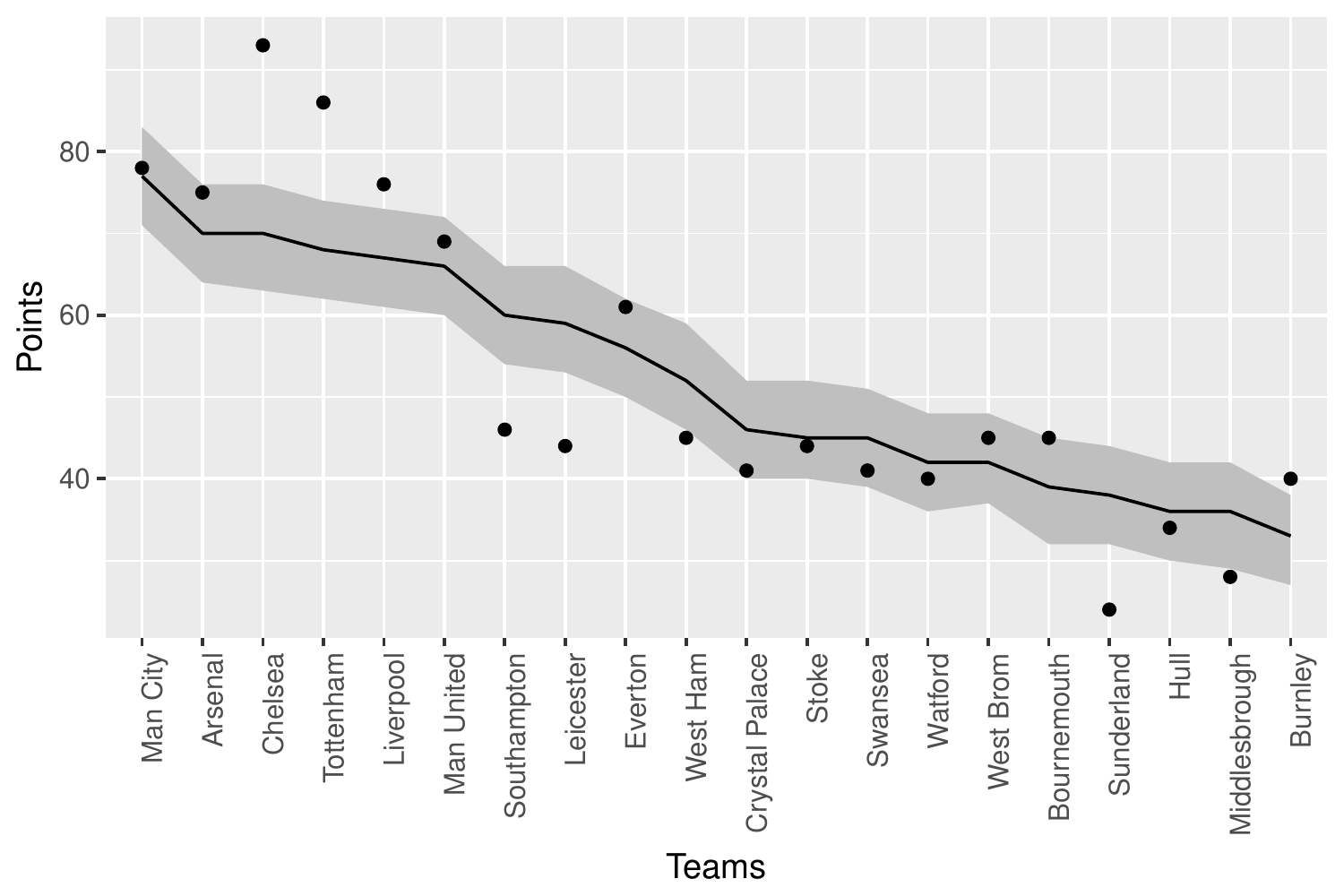}}\\
\subfloat[La Liga]
{\includegraphics[scale=0.5]{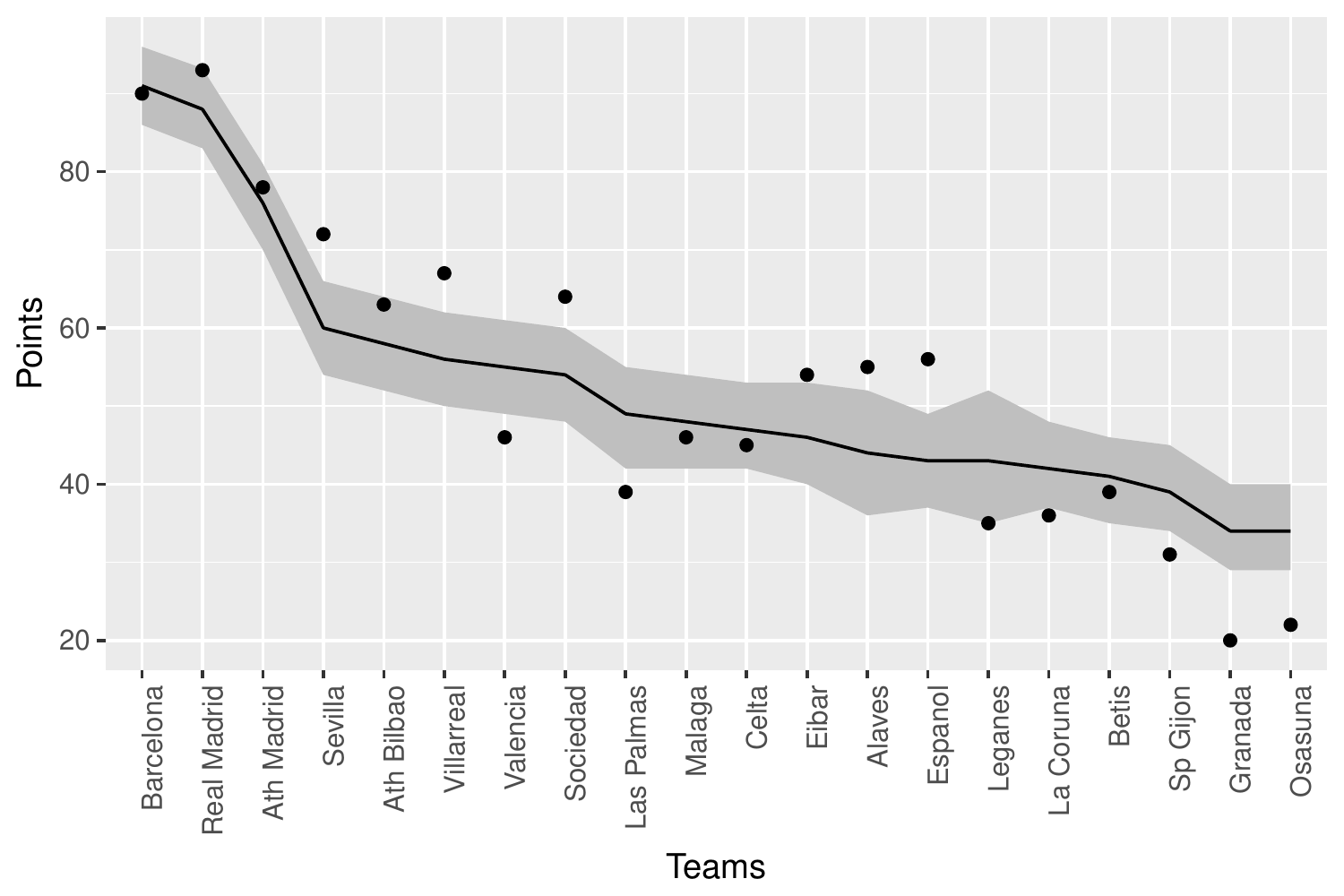}}~
\subfloat[Serie A]
{\includegraphics[scale=0.5]{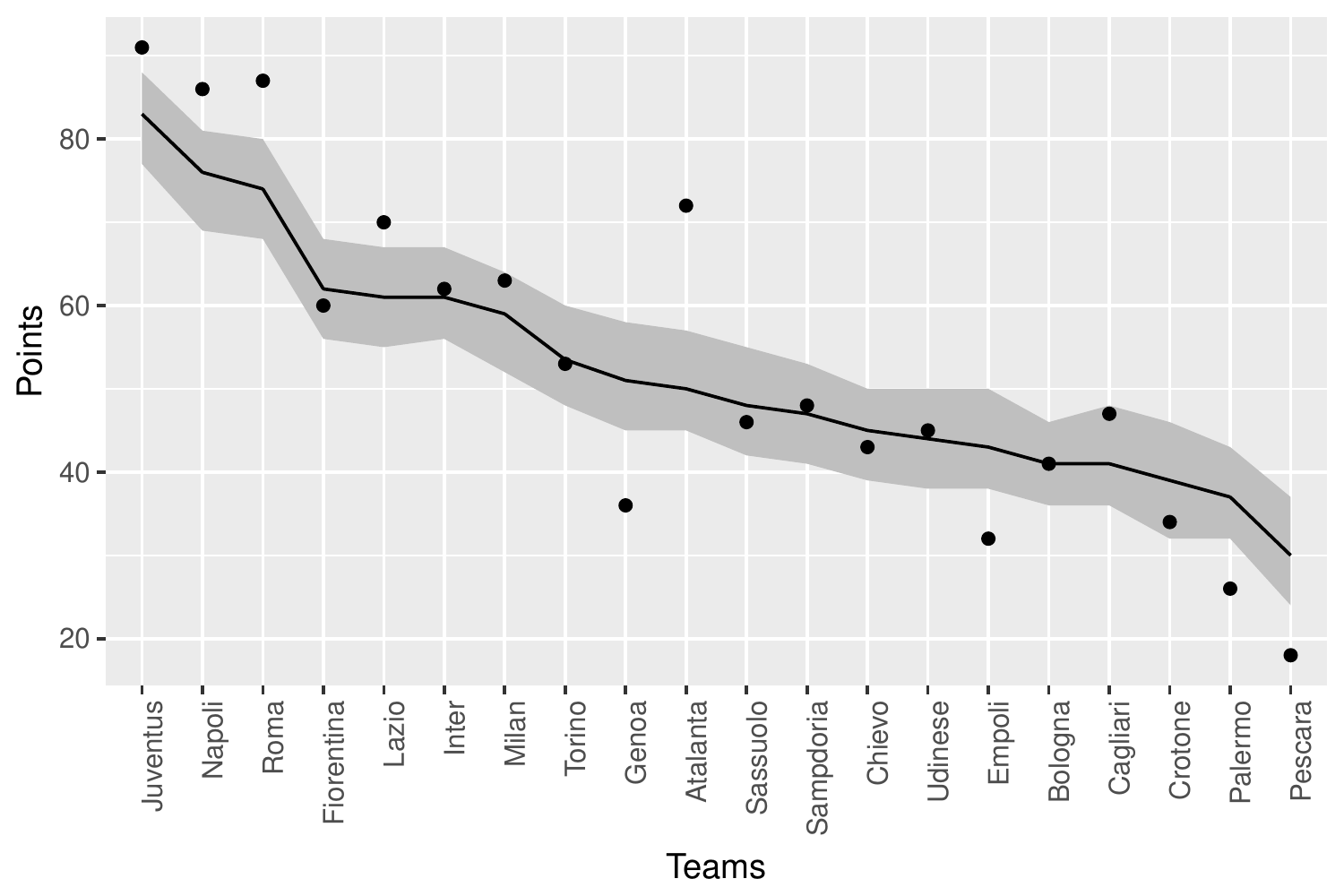}}
\caption{\label{fig04} Posterior 50\% confidence bars (grey ribbons) for the achieved final  points of the top-four European leagues 2016-2017. Black dots are the observed points. Black lines are the posterior medians. At a first glance, the pattern of the predicted ranks appears to match the pattern of the observed ones, and the model calibration appears satisfying.}
\end{figure}

Figure~\ref{fig04} provides posterior 50\% confidence bars (grey ribbons) for the predicted achieved points for each team in top four European leagues 2016-2017 at the end of their 
respective seasons, together with the observed final ranks. At a first glance, the four predicted posterior ranks appear to 
detect a pattern similar to the observed ones, with only a few exceptions. As may be noticed for Bundesliga (Panel (a)), 
Bayern Munich's prediction mirrors its actual strength in the 2016-2017 season, whereas RB Leipzig was definitely 
underestimated by the model. Still, the model cannot handle the budget's information, and RB Leipzig was one of the 
richest teams in the Bundesliga in 2016-2017. In the English Premier League (Panel (b)), Chelsea was definitely 
underestimated by the model, whereas Manchester City actually gained the predicted number of points (78). The predicted 
pattern for the Spanish La Liga (Panel (c)) is extremely close to the one we observed, apart from the winner (our model 
favoured Barcelona, second in the observed rank). The worst teams (Sporting Gijon, Osasuna and Granada) are correctly 
predicted to be relegated. Also, for the Italian Serie A, the predicted ranks globally match the observed ranks. The outlier 
is represented by Atalanta, a team that performed incredibly well and qualified for the Europa League at the end of the last season. As a general comment, we may conclude that these plots show a good model calibration, since more or less 
half of the observed points fall in the posterior 50\% confidence bars.

%\subsubsection{ Results: dynamic prediction day by day}

\section{A preliminary betting strategy}
\label{sec:betting}

In this section we provide a real betting experiment, assessing the performance of our model compared to the 
existing betting odds. In a betting strategy, two main questions arise: it is worth betting on a given single match? 
If so, how much is worth betting? In Section~\ref{sec:elicited}, we described two different procedures for 
inferring a vector of betting probabilities $\Pi$ from the inverse odds vector $O$. The common expression `beating the 
bookmakers' may be interpreted in two distinct ways: from a probabilistic point of view, and from a profitable point of 
view. According to the  first definition, which is more appealing for statisticians, a bookmaker is beaten whenever 
our matches' probabilities are more favorable than their probabilities. As before, $\pi^{s}_{i,m} $ denotes the 
betting probability provided by the $s$-th bookmaker for the $m$-th game, with $i \in \Delta_{m}= \{ \mbox{`Win'}, 
\mbox{`Draw'}, \mbox{`Loss'} \}$. Additionally, let $Y_{m1}$ and $Y_{m2}$ denote the random variables representing 
the number of goals scored by two teams in the $m$-th match. From our model in~\eqref{y:mixture}, we can compute the 
following three-way model's posterior probabilities: ${p}_{Win,m}=P(Y_{m1}>Y_{m2}), \  {p}_{Draw, m}=P(Y_{m1}=Y_{m2}), 
\  {p}_{Loss, m}=P(Y_{m1}<Y_{m2})$ for each $m \in \mathcal{T}_{s}$, using the results of the Skellam distribution outlined 
in Section~\ref{sec:model}. In fact, $Y_{m1}-Y_{m2} \sim PD( \hat{\gamma}_{m1}, \hat{\gamma}_{m2})$, where $\hat{\gamma}
_{m1}=\hat{p}_{m1}\hat{\theta}_{m1}+(1-\hat{p}_{m1})\hat{\lambda}_{m1}$ and $\hat{\gamma}_{m2}=\hat{p}_{m2}
\hat{\theta}_{m2}+(1-\hat{p}_{m2})\hat{\lambda}_{m2}$ are the convex combinations of the posterior estimates obtained 
through the MCMC sampling. Thus, the global average probability of a correct prediction for our model may be defined as: 

\begin{equation}
\bar{p}=\frac{1}{M}\sum_{m=1}^{M} \prod_{i \in  \Delta_{m}} {{p}_{i,m}}^{\delta_{im }},
\label{eq:correct_prob}
\end{equation}
where $\delta_{im}$ denotes the Kronecker's delta, with $\delta_{im}=1$ if the observed result at the $m$-th match is $i, 
\ i \in \Delta_{m}$.  This quantity serves as a global measure of performance for comparing the predictive accuracy between the 
posterior match probabilities provided by the model and those obtained from the bookmakers' odds.
\begin{table}
\caption{\label{tab03} Average correct probabilities $\bar{p}$ of three-way bets, obtained through our model, Shin probabilities and basic probabilities (here we take the average of the seven bookmakers considered). Greater values indicate better predictive accuracy.}
\centering
\centering
\bgroup
\def\arraystretch{0.7}
\begin{tabular}{llccc}
& & Model & Shin & Basic \\
 \hline
\includegraphics[scale=0.05]{deutche.png} & Bundesliga & 0.4010 & 0.4100  &0.4072 \\
\includegraphics[scale=0.05]{inglese.png} & Premier League & 0.4349 & 0.4516 & 0.4480 \\
 \includegraphics[scale=0.05]{spagnola.png}&La Liga & 0.4553 & 0.4584 & 0.4549 \\
 \includegraphics[scale=0.05]{italiana.png}&Serie A & 0.4430 & 0.4554 &0.4507\\
 \hline
\end{tabular}
\egroup
\end{table}
 As reported in Table~\ref{tab03}, our model is very close to the bookmakers' probabilities (Shin's method and basic procedure). At a first 
glance, one may be tempted to say that, according to this measure, our model does not improve the bookmakers' probabilities. However, this index is only an average measure 
of the predictive power, which does not take into account the possible profits for the single matches.

\begin{figure}
\subfloat[Bundesliga]
{\includegraphics[scale=0.45]{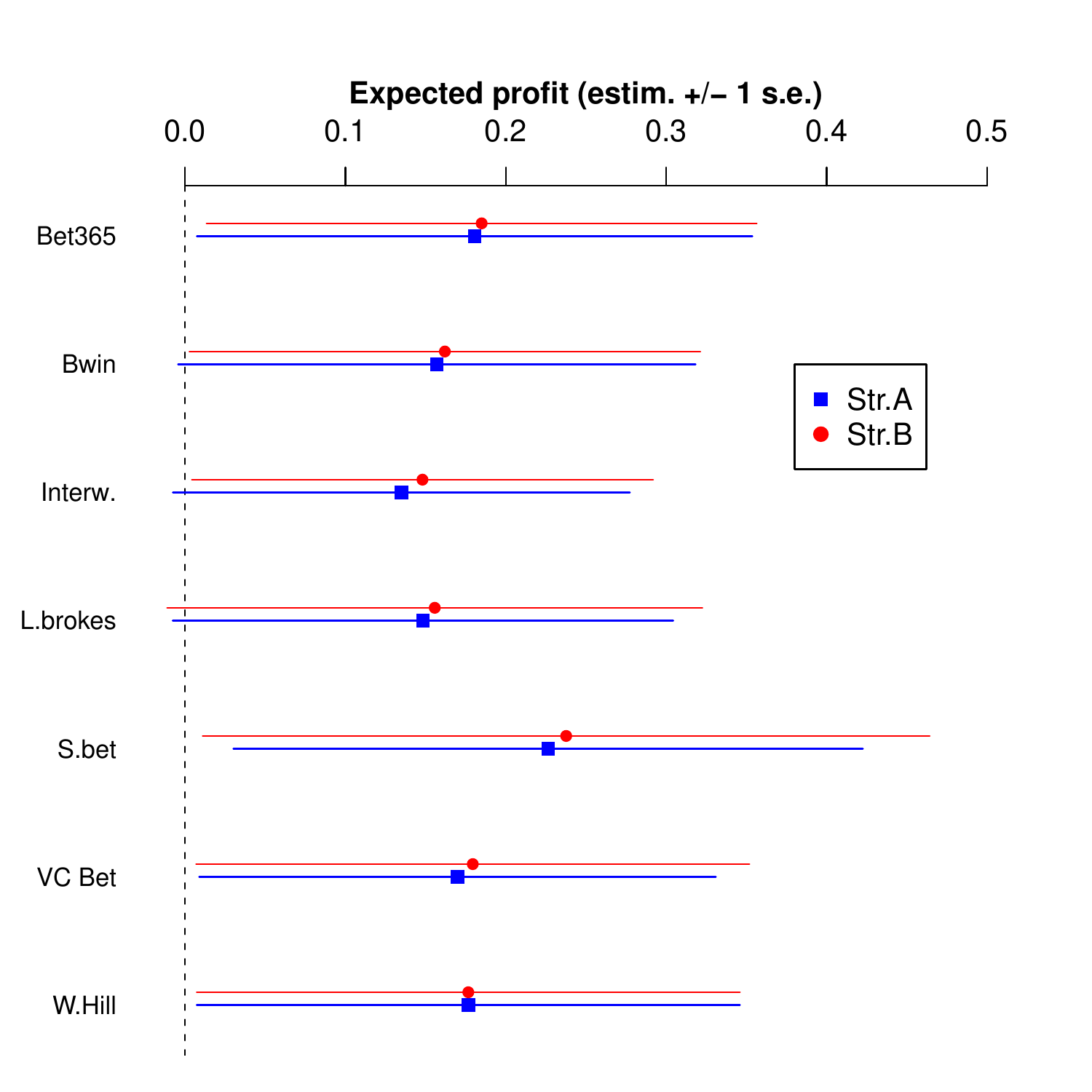}}~
\subfloat[Liga]
{\includegraphics[scale=0.45]{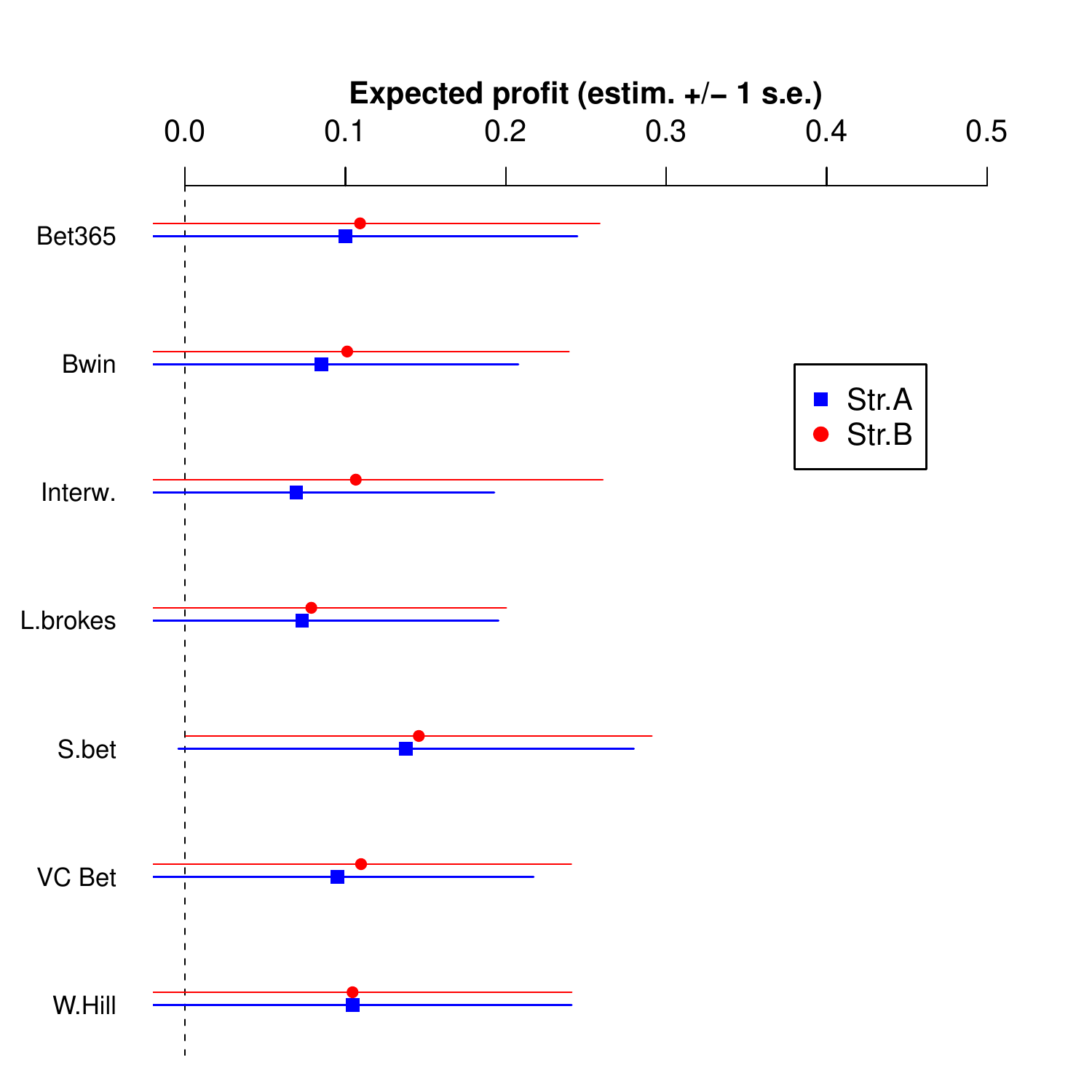}}\\
\subfloat[Premier League]
{\includegraphics[scale=0.45]{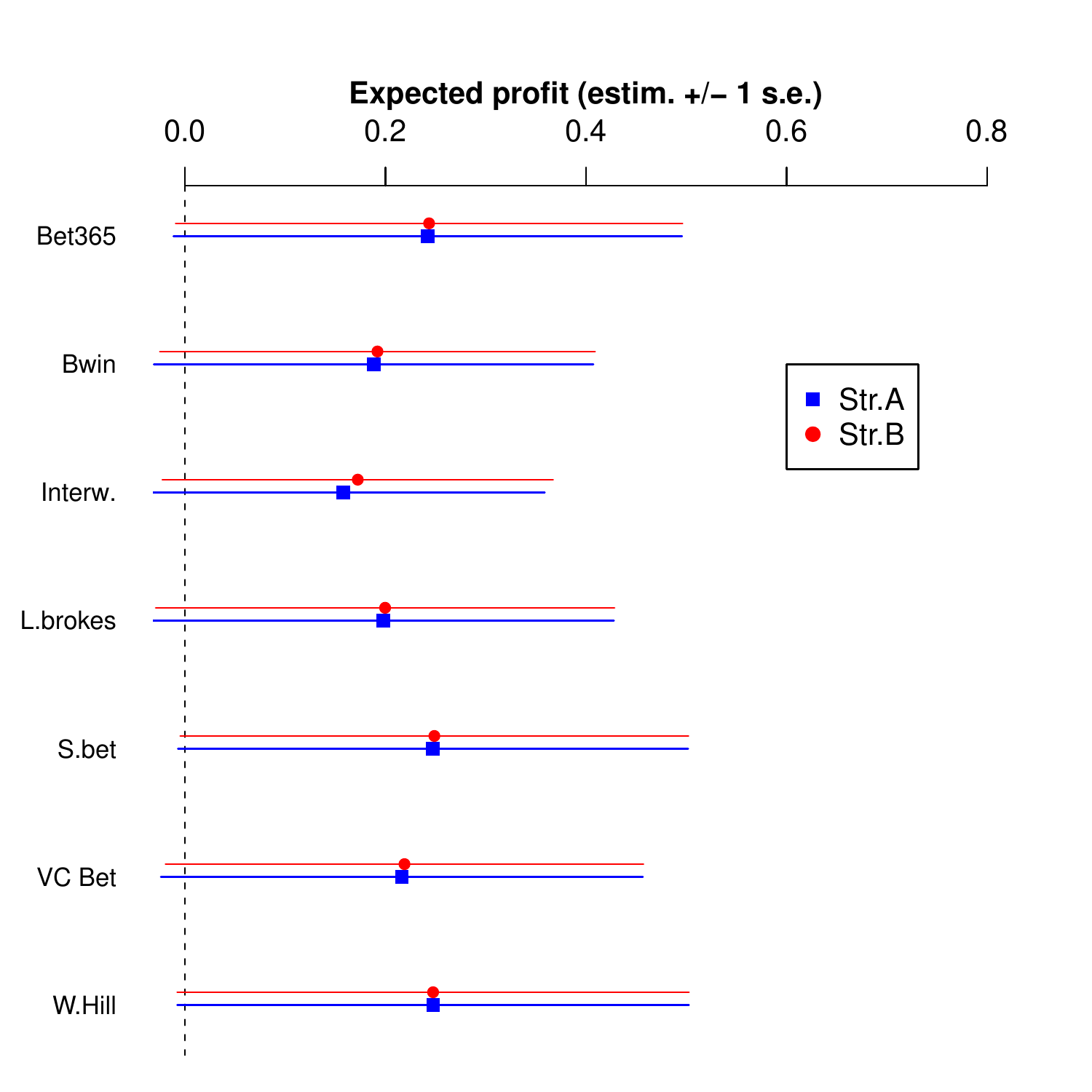}}~
\subfloat[Serie A]
{\includegraphics[scale=0.45]{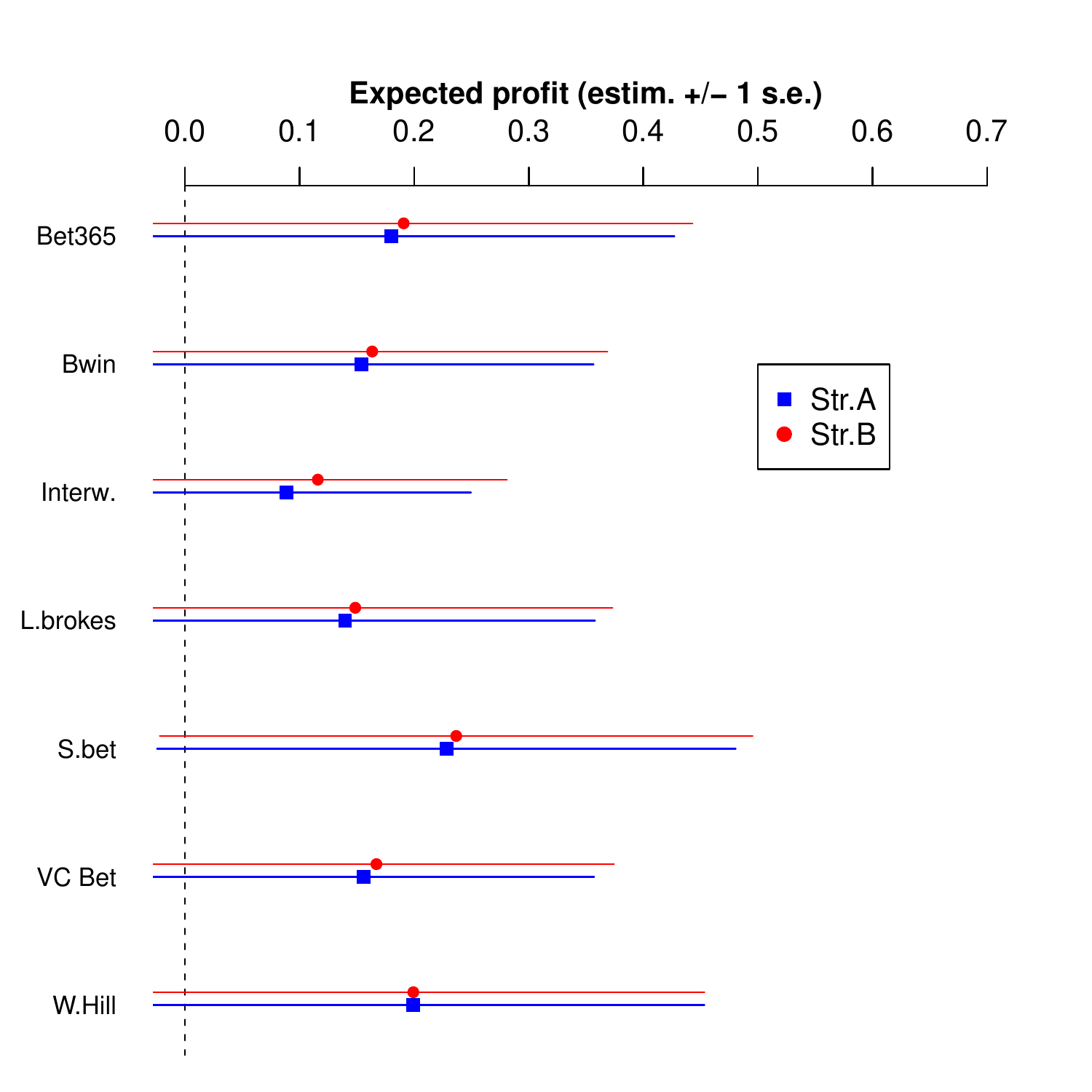}}
\caption{\label{fig_profits} Expected profits (\%/100) $\pm$ standard errors for the seven bookmakers considered, for each of the top four European leagues.  }
\end{figure}

According to the second definition, `beating the bookmaker' means earning money by betting according to our model's probabilities. One 
could bet one unit on the three-way match outcome with the highest expected return (Strategy A) or place different 
amounts, basing each bet on the match's profit variability, as suggested in \cite{rue2000prediction} (Strategy B). The 
expected profits (percentages divided by 100) are reported in Figure~\ref{fig_profits}, along with their standard errors. Although not explicitly 
shown here, gambling with the betting odds probabilities, we would always incur a sure loss. Conversely, betting with 
our posterior model probabilities yields high positive returns for each league and each bookmaker.

\section{Discussion and further work}
\label{sec:concl}

We have proposed a new hierarchical Bayesian Poisson model in which the rates are convex combinations of parameters 
accounting for two different sources of data: the bookmakers' betting odds and the historical match results. We transformed 
the inverse betting odds into probabilities and we worked out the bookmakers' scoring rates through the Skellam distribution. A wide graphical and numerical analysis for the top four European leagues
has shown a good predictive accuracy for our model, and surprising results in terms of expected profits. These results 
confirm on one hand that the information contained in the betting odds is relevant in terms of football prediction; on 
the other hand that, combining this information with historical data allows for a natural extension of the existing models for football scores.   

\bibliography{thesis}

\begin{thebibliography}{}

\bibitem[\protect\citeauthoryear{Baio and Blangiardo}{Baio and
  Blangiardo}{2010}]{baio2010bayesian}
Baio, G. and M.~Blangiardo (2010).
\newblock Bayesian hierarchical model for the prediction of football results.
\newblock {\em Journal of Applied Statistics\/}~{\em 37\/}(2), 253--264.

\bibitem[\protect\citeauthoryear{Cain, Law, and Peel}{Cain
  et~al.}{2002}]{cain2002one}
Cain, M., D.~Law, and D.~Peel (2002).
\newblock Is one price enough to value a state-contingent asset correctly?
  {E}vidence from a gambling market.
\newblock {\em Applied Financial Economics\/}~{\em 12\/}(1), 33--38.

\bibitem[\protect\citeauthoryear{Cain, Law, and Peel}{Cain
  et~al.}{2003}]{cain2003favourite}
Cain, M., D.~Law, and D.~Peel (2003).
\newblock The favourite-longshot bias, bookmaker margins and insider trading in
  a variety of betting markets.
\newblock {\em Bulletin of Economic Research\/}~{\em 55\/}(3), 263--273.

\bibitem[\protect\citeauthoryear{Dixon and Robinson}{Dixon and
  Robinson}{1998}]{dixon1998birth}
Dixon, M. and M.~Robinson (1998).
\newblock A birth process model for association football matches.
\newblock {\em Journal of the Royal Statistical Society: Series D (The
  Statistician)\/}~{\em 47\/}(3), 523--538.

\bibitem[\protect\citeauthoryear{Dixon and Coles}{Dixon and
  Coles}{1997}]{dixon1997modelling}
Dixon, M.~J. and S.~G. Coles (1997).
\newblock Modelling association football scores and inefficiencies in the
  football betting market.
\newblock {\em Journal of the Royal Statistical Society: Series C (Applied
  Statistics)\/}~{\em 46\/}(2), 265--280.

\bibitem[\protect\citeauthoryear{Epstein}{Epstein}{1969}]{epstein1969scoring}
Epstein, E.~S. (1969).
\newblock A scoring system for probability forecasts of ranked categories.
\newblock {\em Journal of Applied Meteorology\/}~{\em 8\/}(6), 985--987.

\bibitem[\protect\citeauthoryear{Forrest, Goddard, and Simmons}{Forrest
  et~al.}{2005}]{forrest2005odds}
Forrest, D., J.~Goddard, and R.~Simmons (2005).
\newblock Odds-setters as forecasters: The case of english football.
\newblock {\em International journal of forecasting\/}~{\em 21\/}(3), 551--564.

\bibitem[\protect\citeauthoryear{Forrest and Simmons}{Forrest and
  Simmons}{2002}]{forrest2002outcome}
Forrest, D. and R.~Simmons (2002).
\newblock Outcome uncertainty and attendance demand in sport: the case of
  english soccer.
\newblock {\em Journal of the Royal Statistical Society: Series D (The
  Statistician)\/}~{\em 51\/}(2), 229--241.

\bibitem[\protect\citeauthoryear{Gelman et~al.}{Gelman
  et~al.}{2006}]{gelman2006prior}
Gelman, A. et~al. (2006).
\newblock Prior distributions for variance parameters in hierarchical models
  (comment on article by {B}rowne and {D}raper).
\newblock {\em Bayesian analysis\/}~{\em 1\/}(3), 515--534.

\bibitem[\protect\citeauthoryear{Gelman, Carlin, Stern, and Rubin}{Gelman
  et~al.}{2014}]{gelman2014bayesian}
Gelman, A., J.~B. Carlin, H.~S. Stern, and D.~B. Rubin (2014).
\newblock {\em Bayesian data analysis}, Volume~2.
\newblock Chapman \& Hall/CRC Boca Raton, FL, USA.

\bibitem[\protect\citeauthoryear{Groll and Abedieh}{Groll and
  Abedieh}{2013}]{groll2013spain}
Groll, A. and J.~Abedieh (2013).
\newblock Spain retains its title and sets a new record--generalized linear
  mixed models on {E}uropean football championships.
\newblock {\em Journal of Quantitative Analysis in Sports\/}~{\em 9\/}(1),
  51--66.

\bibitem[\protect\citeauthoryear{Groll, Schauberger, and Tutz}{Groll
  et~al.}{2015}]{groll2015prediction}
Groll, A., G.~Schauberger, and G.~Tutz (2015).
\newblock Prediction of major international soccer tournaments based on
  team-specific regularized poisson regression: An application to the fifa
  world cup 2014.
\newblock {\em Journal of Quantitative Analysis in Sports\/}~{\em 11\/}(2),
  97--115.

\bibitem[\protect\citeauthoryear{Jullien, Salani{\'e}, et~al.}{Jullien
  et~al.}{1994}]{jullien1994measuring}
Jullien, B., B.~Salani{\'e}, et~al. (1994).
\newblock Measuring the incidence of insider trading: a comment on {S}hin.
\newblock {\em Economic Journal\/}~{\em 104\/}(427), 1418--19.

\bibitem[\protect\citeauthoryear{Karlis and Ntzoufras}{Karlis and
  Ntzoufras}{2003}]{karlis2003analysis}
Karlis, D. and I.~Ntzoufras (2003).
\newblock Analysis of sports data by using bivariate {P}oisson models.
\newblock {\em Journal of the Royal Statistical Society: Series D (The
  Statistician)\/}~{\em 52\/}(3), 381--393.

\bibitem[\protect\citeauthoryear{Karlis and Ntzoufras}{Karlis and
  Ntzoufras}{2009}]{karlis2009bayesian}
Karlis, D. and I.~Ntzoufras (2009).
\newblock Bayesian modelling of football outcomes: using the {S}kellam's
  distribution for the goal difference.
\newblock {\em IMA Journal of Management Mathematics\/}~{\em 20\/}(2),
  133--145.

\bibitem[\protect\citeauthoryear{Koopman and Lit}{Koopman and
  Lit}{2015}]{koopman2015dynamic}
Koopman, S.~J. and R.~Lit (2015).
\newblock A dynamic bivariate poisson model for analysing and forecasting match
  results in the english premier league.
\newblock {\em Journal of the Royal Statistical Society: Series A (Statistics
  in Society)\/}~{\em 178\/}(1), 167--186.

\bibitem[\protect\citeauthoryear{Koopman, Lit, et~al.}{Koopman
  et~al.}{2017}]{koopman2017forecasting}
Koopman, S. J.~S., R.~Lit, et~al. (2017).
\newblock Forecasting football match results in national league competitions
  using score-driven time series models.
\newblock Technical report, Tinbergen Institute.

\bibitem[\protect\citeauthoryear{Londono and Hassan}{Londono and
  Hassan}{2015}]{londono2015sports}
Londono, M.~G. and A.~R. Hassan (2015).
\newblock Sports betting odds: a source for empirical {B}ayes.
\newblock Technical report, EAFIT University, Medellın, Colombia.

\bibitem[\protect\citeauthoryear{Maher}{Maher}{1982}]{maher1982modelling}
Maher, M.~J. (1982).
\newblock Modelling association football scores.
\newblock {\em Statistica Neerlandica\/}~{\em 36\/}(3), 109--118.

\bibitem[\protect\citeauthoryear{McHale and Scarf}{McHale and
  Scarf}{2011}]{mchale2011modelling}
McHale, I. and P.~Scarf (2011).
\newblock Modelling the dependence of goals scored by opposing teams in
  international soccer matches.
\newblock {\em Statistical Modelling\/}~{\em 11\/}(3), 219--236.

\bibitem[\protect\citeauthoryear{Ntzoufras}{Ntzoufras}{2011}]{ntzoufras2011bayesian}
Ntzoufras, I. (2011).
\newblock {\em Bayesian modeling using WinBUGS}, Volume 698.
\newblock John Wiley \& Sons.

\bibitem[\protect\citeauthoryear{Owen}{Owen}{2011}]{owen2011dynamic}
Owen, A. (2011).
\newblock Dynamic bayesian forecasting models of football match outcomes with
  estimation of the evolution variance parameter.
\newblock {\em IMA Journal of Management Mathematics\/}~{\em 22\/}(2), 99--113.

\bibitem[\protect\citeauthoryear{Rue and Salvesen}{Rue and
  Salvesen}{2000}]{rue2000prediction}
Rue, H. and O.~Salvesen (2000).
\newblock Prediction and retrospective analysis of soccer matches in a league.
\newblock {\em Journal of the Royal Statistical Society: Series D (The
  Statistician)\/}~{\em 49\/}(3), 399--418.

\bibitem[\protect\citeauthoryear{Shin}{Shin}{1991}]{shin1991optimal}
Shin, H.~S. (1991).
\newblock Optimal betting odds against insider traders.
\newblock {\em The Economic Journal\/}~{\em 101\/}(408), 1179--1185.

\bibitem[\protect\citeauthoryear{Shin}{Shin}{1993}]{shin1993measuring}
Shin, H.~S. (1993).
\newblock Measuring the incidence of insider trading in a market for
  state-contingent claims.
\newblock {\em The Economic Journal\/}~{\em 103\/}(420), 1141--1153.

\bibitem[\protect\citeauthoryear{Smith, Paton, and Williams}{Smith
  et~al.}{2009}]{smith2009bookmakers}
Smith, M.~A., D.~Paton, and L.~V. Williams (2009).
\newblock Do bookmakers possess superior skills to bettors in predicting
  outcomes?
\newblock {\em Journal of Economic Behavior \& Organization\/}~{\em 71\/}(2),
  539--549.

\bibitem[\protect\citeauthoryear{Spiegelhalter, Thomas, Best, and
  Lunn}{Spiegelhalter et~al.}{2003}]{spiegelhalter2003winbugs}
Spiegelhalter, D., A.~Thomas, N.~Best, and D.~Lunn (2003).
\newblock {\em WinBUGS user manual}.

\bibitem[\protect\citeauthoryear{{Stan Development Team}}{{Stan Development
  Team}}{2016}]{rstan}
{Stan Development Team} (2016).
\newblock {RStan}: the {R} interface to {S}tan, version 2.9.0.

\bibitem[\protect\citeauthoryear{{\v{S}}trumbelj}{{\v{S}}trumbelj}{2014}]{vstrumbelj2014determining}
{\v{S}}trumbelj, E. (2014).
\newblock On determining probability forecasts from betting odds.
\newblock {\em International journal of forecasting\/}~{\em 30\/}(4), 934--943.

\end{thebibliography}

\end{document}